\documentclass[sn-mathphys]{sn-jnl}

\jyear{2021}%

\theoremstyle{thmstyleone}%
\theoremstyle{thmstyletwo}%

\theoremstyle{thmstylethree}%

\usepackage{amsfonts}
\usepackage{amsmath}
\usepackage{amssymb}
\usepackage{cleveref}

\def\dr{\overline {\rm DR}}
\def\t{\tilde }

\newcommand{\jl}[1]{{\color{blue} \bf{#1}}}

\newcommand{\higgsmolar}{\ensuremath{\mathrm{HiggsMolar}}}
\newcommand{\cubicFP}{\ensuremath{\mathrm{C_{FP}}}}
\newcommand{\linearFP}{\ensuremath{\mathrm{L_{FP}}}}

\begin{document}

\title[The Higgs Mass 
as Parameter of the MSSM]{The Higgs Boson Mass 
as Fundamental Parameter of the Minimal Supersymmetric Standard Model}


\author[1,5]{\fnm{Rima} \sur{El-Kosseifi}}

\author[2]{\fnm{Jean-Loic} \sur{Kneur}}\email{jean-loic.kneur@umontpellier.fr}

\author*[2]{\fnm{Gilbert} \sur{Moultaka}}\email{gilbert.moultaka@umontpellier.fr}

\author[3,4]{\fnm{Dirk} \sur{Zerwas}}\email{Dirk.Zerwas@in2p3.fr}

\affil[1]{\orgdiv{CPPM}, \orgname{Aix-Marseille Universit\'e, CNRS/IN2P3}, \orgaddress{\city{Marseille}, \country{France}}}

\affil[2]{\orgdiv{Laboratoire Charles Coulomb (L2C)}, \orgname{ Universit\'e de Montpellier, CNRS}, \orgaddress{\city{Montpellier}, \country{France}}}

\affil[3]{\orgdiv{IJCLab}, \orgname{Universit\'e Paris-Saclay, CNRS/IN2P3}, \orgaddress{\city{Orsay}, \country{France}}}

\affil[4]{\orgdiv{DMLab}, \orgname{CNRS/IN2P3}, \orgaddress{\city{Hamburg}, \country{Germany}}}

\affil[5]{Now at Apneal/Mitral SAS, \city{Charenton-le-Pont}, \country{France}}

\abstract{In the Minimal Supersymmetric Standard Model (MSSM) the mass of the lightest neutral Higgs boson is determined by the supersymmetric parameters. In the $m_h$MSSM the precisely measured Higgs boson replaces the trilinear coupling $A_t$ as input parameter. Expressions are derived to extract $A_t$ in a semi-analytical form as a function of the light 
Higgs boson (pole) mass. An algorithm is developed and implemented at two--loop precision,
generalizable to higher orders, to perform this inversion consistently. 
The result of the algorithm, implemented in the SuSpect spectrum calculator, is illustrated on a parameter set compatible with LHC measurements.}




\maketitle

\section{Introduction}

In the Minimal Supersymmetric Standard Model (MSSM) the scalar boson discovered by ATLAS and CMS~\cite{Aad:2012tfa,Chatrchyan:2012ufa} is identified with the lightest 
neutral Higgs boson of the model. Its mass has been determined to be 125.10~GeV with a precision of 0.14~GeV when combining the measurements of ATLAS and CMS~\cite{Aaboud:2018wps,Sirunyan:2017exp,Aad:2015zhl,ParticleDataGroup:2020ssz}.

Given the precision of the measurement it is tempting 
to express model parameters of the MSSM as a function of this measurement, as it is customarily done for the fermion masses. 
This choice is also similar to e.g, the almost universally adopted convention of expressing 
electroweak BSM model parameters as functions of the Z-boson pole mass input, $m_Z$, after its 
precise determination at LEP1 in the early 1990's~\cite{ALEPH:2005ab}. This procedure was 
analytically nontrivial beyond tree-level as it necessarily involved 
the radiative corrections 
contributing to the Z-boson pole mass (for the state of the art see~\cite{Dubovyk:2019szj} and references therein). 
In the MSSM, the relation of the lighest Higgs boson mass $m_h$ to the basic model
parameters is even more involved when including the 
radiative corrections. 

The inversion of relations between parameters of a model and a physical observable is
facilitated by approximations which are more easily amenable to such a procedure. 
In the MSSM this approach has been studied in the gaugino-Higgsino 
sector~\cite{Kneur:1998gy,Kneur:1999nx,Kneur:2008ur} and in the Higgs sector~\cite{Kneur:2008ur,Djouadi:2013uqa}. By construction the precision of the approach depends on the precision of the approximation which has to be compared to the precision of the experimental
measurement. Thus the development of an algorithm to cope with the highest available precision is developed in this paper. 

There is a second motivation for the replacement of the model parameter. If a full exploration 
of the MSSM parameter space is performed, a large fraction of the parameter sets studied will 
not predict a lightest neutral Higgs boson mass in agreement with the experimental measurement. Using $m_h$ as parameter has the potential to lead to a more efficient exploration of the MSSM. 

In the following the MSSM with $m_h$ as parameter is referred to as $m_h$MSSM to differentiate the model 
from the standard MSSM. It is well known that $m_h$
depends non-trivially and strongly on the trilinear coupling $A_t$ as well 
as on the third generation squark sector soft breaking masses via its leading radiative corrections. 
Because of the $A_t \leftrightarrow m_h$ connection, it is natural to choose to develop the inversion with $A_t$ replaced by $m_h$ 
as model parameter in the $m_h$MSSM. The study is carried out within the CP-conserving version of the MSSM where $A_t$ is real-valued, and relies on the fixed-order 
loop approximations. We will comment on departure from these assumptions towards the end of the paper.

Inverting a relationship between two parameters can be performed in multiple ways. A brute force approach would be
to calculate $m_h$ as fonction of all MSSM parameters as well as their SM inputs, taking into account the experimental
and systematic errors. This is time consuming and inefficient as in most of the parameter space the predicted $m_h$ is too light with respect to the experimental measurement. Machine Learning algorithms are more efficient but need extensive training
and validation. For each update of the calculations the full procedure of determination and validation has to be performed again. For these reasons an algorithmically simple procedure, the fixed point algorithm, is used. It has the advantage that its
preparation is analytical work, i.e., an appropriate function has to be derived and the convergence criteria have to be fulfilled. The guiding principle behind this choice is that the additional calculations should add minimal overhead to the calculation of the spectrum.

The paper is organized as follows. In \cref{sec:MSSMHmass} the basic generic expressions
for the scalar Higgs boson masses in the MSSM, including higher order radiative corrections, 
are recalled and the notations are defined. The MSSM parameter set to serve as test case for our
general approach is defined.
In \cref{sec:HtoAt}  the inversion $A_t(m_h)$ is
first illustrated  
using an analytical approximation at one--loop level, whose purpose is to serve as 
first guess within a subsequent more elaborate construction. Then the exact full one--loop expression is given together with a description of the strategy for the inversion algorithm, as well as the consistent implementation of the 
dominant two--loop contributions.  
In \cref{sec:Algo} the full algorithm is assembled by combining the analytical approximation with the full one--loop and
dominant two--loop calculations. The results of its application to the parameter set are discussed, as well as possible extensions. We conclude in \cref{sec:conclusion} and provide technical material
in the appendices.

\section{Setting up the MSSM Higgs Mass \label{sec:MSSMHmass}}

In this section we briefly recall the main content of the 
fixed-order (diagrammatic) calculation of $m_h$. Then we define a parameter set to illustrate numerical results.

\subsection{Calculation of the Higgs Boson Mass \label{sec:HiggsMass}}

In the standard 'top-down' procedure one assumes that all the SM and MSSM parameters
are taken as input before determining the light (and heavy) CP-even Higgs boson masses through the diagonalization of the corresponding (momentum dependent) squared mass matrix.  The latter has the following generic form:

\begin{align}
&{\cal M}^2_s(p^{2})\! = \!\!
\left(\!\!\begin{array}{ll}
\displaystyle \overline{m}_{11}^2-\Pi_{11}(p^{2})+\frac{t_1}{v_1} &  
\overline{m}_{12}^2- \Pi_{12}(p^{2})\\ 
\overline{m}_{12}^2- \Pi_{12}(p^{2})
 &  \displaystyle \overline{m}_{22}^2 -\Pi_{22}(p^{2})+\frac{t_2}{v_2}
\end{array}\!\!\right)
\label{eq:Ms}
\end{align}
where 
\begin{align}
&\overline{m}_{11}^2 =\overline{m}_{Z}^{2} \cos^{2}\beta +  \overline{m}_{A}^{2}  \sin^{2}\beta, \label{eq:m112}\\
&\overline{m}_{22}^2 =\overline{m}_{Z}^{2} \sin^{2}\beta +  \overline{m}_{A}^{2}  \cos^{2}\beta, \label{eq:m222}\\
&\overline{m}_{12}^2 =-\frac{1}{2}(\overline{m}_{Z}^{2} +  \overline{m}_{A}^{2})  \sin 2\beta. \label{eq:m122}
\end{align}
Here $\overline{m}_{Z}^{2}$ and $\overline{m}_{A}^{2}$ denote
the running $Z$-boson and CP-odd Higgs boson squared masses, $\beta$ the angle defined by $\tan\beta=v_2/v_1$ where $v_1,v_2$ are the two Higgs vacuum expectation values, and 
$\Pi_{ij}(p^{2})$ and $t_i$ designate respectively the renormalized self-energy and tadpole loop contributions, formally to arbitrary orders in perturbation theory.
In practice we rely on the $\overline{DR}$ renormalization scheme.
For reviews on radiative corrections to $m_h$ in the MSSM and original references see
e.g. \cite{Djouadi:2005gj,Heinemeyer:2004ms}, and 
\cite{Slavich:2020zjv} for a recent up--to-date review.

In \cref{eq:Ms} the tree-level contributions involve $\overline{m}_{A}$
evaluated at a given electroweak symmetry breaking (EWSB) scale $Q_{EWSB}$, and the MSSM parameter $\tan\beta$. 
If we ignore momentarily the complication that the running parameters in the expressions above have actually 
a non-trivial implicit dependence on the other MSSM parameters, 
the ''tree-level``  masses of the two CP-even MSSM Higgs states and their 
mixings are described by only these two unknown MSSM parameters as well as $\overline{m}_Z$. 

Beyond the tree-level the squared mass matrix depends on the external squared momentum $p^2$ through the self-energies, as shown in \cref{eq:Ms}. The actual {\em pole} masses, $m_h$ and $m_H$, 
are then obtained by determining the two solutions $p^2=p_h^2$ and  $p^2=p_H^2$ to the equation
\begin{equation}
\label{eq:mhpole}
\det \left(p^2 {\mathbf 1} - {\cal M}^2_s(p^{2}) \right)=0,
\end{equation}
and taking $m_{h,H}=\sqrt{\operatorname{Re}( p_{h,H}^2)}$.  
Obviously, the loop contributions in ${\cal M}^2_s(p^{2})$ 
depend also on the chosen ($\overline{DR}$) renormalization scale. 

In this paper we focus on the lighter CP-even Higgs with mass $m_h$.
The self-energies and tadpoles contained in ${\cal M}^2_s(p^{2})$ have been known to one--loop order exactly since the 1990's~\cite{Chankowski:1991md,Chankowski:1992er,Dabelstein:1994hb,Pierce:1996zz},  
as well as the QCD\cite{Hempfling:1993qq,Heinemeyer:1998jw} and other dominant two-loop corrections in the on-shell 
scheme~\cite{Heinemeyer:1998np,Heinemeyer:1998kz,Heinemeyer:2004xw}
 or in the $\dr$ scheme~\cite{Zhang:1998bm,Espinosa:1999zm,Espinosa:2000df,Degrassi:2001yf,
 Brignole:2001jy,Brignole:2002bz,Dedes:2002dy,Dedes:2003km,Degrassi:2002fi}.
The (almost) complete two--loop contributions are also available~\cite{Martin:2002wn,Martin:2004kr,Borowka:2014wla,Degrassi:2014pfa}, as well as
the dominant higher order contributions~\cite{Martin:2007pg,Harlander:2008ju,Kant:2010tf,Harlander:2017kuc}. These have been included in some analyses (see \cite{Slavich:2020zjv} for details). 
In practice, the determination of the pole mass $m_h$  
is achieved by iterating on its implicit 
expression \cref{eq:mhpole}, until a sufficient accuracy is reached.

Let us now recall some important features of the scalar sector parameter relationship that will be relevant for our construction. Depending on the phenomenological context, $\overline{m}_{A}$ may be either an input or a derived quantity. 
In the first case, typical for low-energy model-independent applications, $\overline{m}_{A}$ can be a direct input at a given EWSB scale $Q_{EWSB}$, or inferred from the {\sl pole} mass $m_A$ taken as input. In the second case, typical 
for top-down approaches, 
it is obtained from the Supersymmetry (SUSY) soft–breaking running Higgs mass parameters  $\overline{m}_{H_{u}}$ and $\overline{m}_{H_{d}}$,
evolved by the renormalization group equations (RGE) down to a scale $Q_{EWSB}$ where the EWSB constraints
are imposed:

\begin{align}
 \label{eq:MAEWSB}
\overline{m}_{A}^{2}(Q_{EWSB})&=   \frac{1}{\cos 2 \beta}\left(\hat{m}_{H_{u}}^{2}-\hat{m}_{H_{d}}^{2}\right)-\overline{m}_Z^2 , 
\\
\label{eq:muEWSB}
\mu^2(Q_{EWSB})&=\frac12 \Big( \left(\hat{m}_{H_{u}}^{2} \tan \beta - \hat{m}_{H_{d}}^{2} \cot \beta \right)\tan 2 \beta  -\overline{m}_Z^2 \Big). 
\end{align}
Here $\mu$ denotes the running supersymmetric Higgs mixing parameter, and $\hat{m}_{H_{i}}^{2} \equiv \overline{m}_{H_{i}}^{2} - t_i/v_i$.

The self-energies and tadpole loop contributions in \cref{eq:Ms} depend implicitly 
on all MSSM parameters through their sensitivity to the couplings and masses of (s)particles
entering the loops. In particular, the dominant radiative correction originates from the top quark mass $m_{t}$ as well as the stop masses and mixing, $m_{\tilde{t}_{i}}$, $X_t\equiv 
A_t-\mu /\tan\beta$. A further implicit dependence on these parameters occurs when Eqs.~(\ref{eq:MAEWSB}) and (\ref{eq:muEWSB}) are imposed. This will be instrumental for the identification of the functional dependence of $m_h$ on the trilinear stop coupling $A_t$, in order to set up an efficient inversion algorithm leading to $A_t(m_h)$, i.e. in the $m_h$MSSM.


\subsection{Stop Cliff}

\label{sec:Points}

\begin{table}[htbp]
\begin{center}
\caption{The Stop Cliff set of MSSM parameters is given. The scale dependent parameters are defined at the EWSB scale with the exception of $\tan\beta$ which is defined at the Z~boson scale. The resulting masses of the Higgs, lightest stop and LSP are listed as well.}\label{tab:MhInvPoints}
\begin{tabular}{lr}
\toprule
EW                  & 2.0 TeV   \\
\midrule
$m^2_{H_d}$          & 3.65740418 TeV$^2$ \\
$m^2_{H_u}$         &  -0.213361994 TeV$^2$  \\
sign($\mu$)         & $+$ \\
\midrule
$A_t$                & 3.610 TeV   \\
$m_{\tilde{t}_R}$    & 1.27 TeV   \\
\midrule
$m_{\tilde{q}3_L}$   & 3 TeV\\
\midrule
$M_1$                & 300 GeV     \\
\midrule
$M_2$                & 2 TeV  \\
$M_3$                & 3 TeV  \\
$A_b$, $A_\tau$      & 0 GeV  \\
$\tan\beta$          & 10 \\
$m_{\tilde{e}_L} =
m_{\tilde{\mu}_L} =
m_{\tilde{\tau}_L} =
m_{\tilde{e}_R} =
m_{\tilde{\mu}_R}= 
m_{\tilde{\tau}_R}$ & 2 TeV  \\
$m_{\tilde{q}1_L}=
m_{\tilde{q}2_L}=
m_{\tilde{u}_R}=
m_{\tilde{c}_R}=
m_{\tilde{d}_R}=
m_{\tilde{s}_R}=
m_{\tilde{b}_R}$    & 3 TeV  \\
\midrule
$m_h$                & 125.012 GeV \\
$m_{{\tilde{t}}_1}$          &  1306 GeV   \\
$m_{{\tilde{\chi}}_1^\circ}$       &   294 GeV   \\
\botrule
\end{tabular}
\end{center}
\end{table}
In order to illustrate the general procedure with a concrete example, an MSSM benchmark parameter set is used in the study. 
The parameters are listed in \cref{tab:MhInvPoints}. 

The EWSB scale is fixed as suggested in \cite{AguilarSaavedra:2005pw}. 
The numerical value chosen is close to the geometric mean of the top squark masses $m_{\tilde{t}_{1}},m_{\tilde{t}_{2}}$. 
In the following we will use as input parameter either $A_t$ in the MSSM 
or alternatively $m_h$ in the $m_h$MSSM. Many parameters
of the table have little influence on the value of $m_h$. They are set to values which are sufficiently large to evade the lower bounds on supersymmetric particle masses
determined at the LHC~\cite{ATLAS:2020syg,ATLAS:2019lff,CMS:2019zmd,CMS:2021cox,CMS:2020bfa}. 

The value of $M_1$ is chosen to obtain a dominantly Bino LSP
$\chi_1^\circ$. Of the two soft breaking masses $m_{\tilde{q}3_L}$ is much greater than $m_{\tilde{t}_R}$, therefore the lightest top squark is essentially of type $R$. $m_{\tilde{t}_R}$ dominates the determination of its
mass as the mixing angle is small. The resulting masses of the lightest top squark and the LSP are shown in \cref{tab:MhInvPoints} as well. The top squark mass of 1.3~TeV was chosen to be close to 
the exclusion bounds determined at the LHC by ATLAS~\cite{ATLAS:2020dsf} and CMS~\cite{CMS:2021beq}.
We will refer to the benchmark point of \cref{tab:MhInvPoints} as the Stop Cliff.

In order to accurately compare the $m_h$MSSM to the MSSM determination, we have first adjusted $A_t$ such that $m_h(A_t)$ is driven to its experimental value. The resulting $m_h$ values are rounded to MeV. This is two orders of magnitude more precise than the current experimental precision.


\section{From the Higgs Boson Mass to the Trilinear Stop Coupling}
\label{sec:HtoAt}

The dependence of
$m_h$ on $A_t$ is non-trivial. 
First a simple well-known approximation
will be studied, followed by a full one--loop expression derivation. Finally the two--loop contributions will be included.

The strategy to determine $A_t(m_h)$ relies on identifying first the
algebraic dominant dependencies on $A_t$ at the given perturbative order or approximation.
Then either the approximation is solved for $A_t$ or the resulting equation
at given order is transformed into a fixed-point problem. The latter,
via the intrinsically iterative structure of the determination, will
account also for the residual non-algebraic dependencies, leading to the
exact determination of $A_t(m_h)$.

\subsection{Approximate One--Loop Inversion}
\label{subsec:AppOneLoop}
Several approximate expressions have been developed in the past for the dominant radiative corrections to $m_h$ in the MSSM,
ranging from simple to sophisticated~\cite{Ellis:1990nz,Haber:1990aw,Okada:1990gg},\cite{Carena:1995bx},\cite{Carena:1995wu,Haber:1996fp,Carena:2000dp,Heinemeyer:2000nz}. While the latter cannot compete with the full one--loop plus two--loop 
calculations available nowadays, some of the approximations including dominant two--loop contributions can be rather precise, depending on the considered MSSM 
parameter range.
In our construction there is no need to rely on the most elaborate approximations.
The specific expression that we will use is 
 the well-known one originally derived in \cite{Carena:1995bx},\cite{Carena:1995wu}, see also \cite{Carena:2000dp}, obtained from considering only the 
dominant top and stop contributions to the one--loop MSSM effective potential, and 
using renormalization group properties to resum the leading logarithms  
of decoupled "heavy" stops (relative to $\sim m_t$):
%
%
\begin{equation}
  \label{eq:AppRC}
 m^2_h = \overline{m}_{h}^{2} +\dfrac{3\overline{g}_{2}^{2}\overline{m}_{t}^{4}}{8 \pi^{2} \overline m_{W}^{2}} 
 \left[\ln\left(\dfrac{M_{S}^{2}}{\overline{m}_{t}^{2}}\right)
 +\dfrac{X_{t}^{2}}{M_{S}^{2}}-\dfrac{X_{t}^{4}}{12 M_{S}^{4}}\right],
 \end{equation}
 with the noteworthy quartic and quadratic dependencies on the stop mixing parameter, 
 \begin{equation}
 X_{t}\equiv A_{t}- \mu \cot\beta .
 \end{equation} 
 In \cref{eq:AppRC}, $\overline{m}_h$ is the (running) tree-level Higgs mass
 \begin{equation}
 \label{eq:mhtree}
\overline{m}_{h}^{2} = \dfrac{1}{2}\left[ 
\overline{m}_{Z}^{2} +  \overline{m}_{A}^{2} -
\sqrt{(\overline{m}_{A}^{2}-\overline{m}_{Z}^{2})^2+
4 \overline{m}_{Z}^{2} \overline{m}_{A}^{2} \sin^2 (2\beta) } \right], 
 \end{equation}
 $\overline{m}_{t}$ the running top mass, $\overline g_{2}$ the running SU(2) gauge coupling, and $ \tan \beta$ the running vev ratio,
and we define
 \begin{align}
 \label{eq:MSusy}
   M_{S}^{2}= & \sqrt{(\overline m_{\tilde{q}3_L}^{2} +(\frac{1}{2}-\frac{2}{3} s_W^2 ) \overline m_Z^2 \cos{2\beta}+ \overline m_t^2)}\cdot
  \sqrt{(\overline m_{\tilde{t}_{R}}^{2} +\frac{2}{3}s_W^2 \overline m_Z^2\cos{2\beta} + \overline m_t^2)}
\end{align}
where $\overline m_{\tilde{q}3_L}^{2}$ and $\overline m_{\tilde{t}_{R}}^{2}$ are the running soft 
SUSY-breaking parameters of the third generation associated to the left doublet and the stop singlet.

We choose to evaluate all the running quantities in \cref{eq:AppRC} at the EWSB scale $Q_{EWSB}\simeq (m_{\tilde{t}_1} m_{\tilde{t}_2})^{1/2}$, including $\overline m_t$. While this choice for $\overline m_t$, and the definition of $M_S$ in \cref{eq:MSusy}, are somewhat at variance with the literature, we emphasize here that we seek a sufficiently 
accurate but simple expression whose sole purpose is to serve as a first guess for our genuine 
algorithm, the latter giving consistently a (perturbatively)
''exact`` $A_t(m_h)$. To put things in perspective, let us enumerate some important features related to \cref{eq:AppRC}, referring to~\cite{Carena:1995bx,Carena:1995wu,Haber:1996fp,Carena:2000dp} for details: 
\begin{itemize}
\item Strictly speaking, \cref{eq:AppRC} is valid for 
large $\tan\beta$, neglecting ${\cal O}(g^4_i)$ terms,  
and in the limit $M_S \gg m_t$ 
as an expansion in 
$\frac{\lvert m_t X_t\rvert}{M_S^2}$. 
In particular, 
among other necessary steps\cite{Carena:1995bx} in the derivation of \cref{eq:AppRC} from the MSSM effective potential,
the $X^4_t$ term 
arises only after expanding to second order in $X_t^2$ the terms with logarithmic dependence in the stop masses, 
$ \ln m^2_{\tilde{t}_i}\simeq 
\ln (M_S^2 \pm \overline m_t X_t)$ (in the limit $M_S\simeq \sqrt{m_{\tilde{q}3_L}^2 + m_t^2} \simeq \sqrt{m_{\tilde{t}_{R}}^2 + m_t^2}$). 
This approximation was thus
a priori expected to be valid only for rather
moderate $\lvert X_t/M_S\rvert$, large $\tan\beta$ and for $m_A \gg m_Z$, but turned out to be reasonably good in an extended range. 

\item In \cite{Carena:1995bx},\cite{Carena:1995wu,Haber:1996fp,Carena:2000dp},
the $D$-terms $\propto m_Z$
in Eq.(\ref{eq:MSusy}) were accordingly neglected, moreover,  universal soft masses,  $m_{\tilde{q}3_L} =m_{\tilde{t}_{R}}\simeq M_{susy}$, were also assumed for simplicity. 
Actually in the effective field theory (EFT) framework 
appropriate to derive Eq.(\ref{eq:AppRC}), 
$M_S$ is not very precisely defined as long as it is 
identified as the scale at which the
top squarks are decoupled, and the matching to the EFT is done: $M_S$ is assumed in the literature  to be of order
the average stop masses. In a more refined treatment (or to account for $m_{\tilde{t}_1} \ll m_{\tilde{t}_2}$) one would need to decouple the two stop masses {\em separately}, 
which is beyond the scope of 
the approximation \cref{eq:AppRC}. 
Our slightly different definition in \cref{eq:MSusy}
makes minor differences, the important practical feature for our purpose being that $M_S$ in Eq.(\ref{eq:MSusy}) does not
involve an extra dependence on $X_t$.
\item The scale $Q_t$ at which  
the running top  mass $\overline{m}_t(Q_t)$ is evaluated in \cref{eq:AppRC} is quite relevant due to the $\overline{m}_t^4$
dependence. While the EFT one--loop calculation in \cite{Carena:1995bx},\cite{Carena:1995wu} 
involves $\overline m_t(m_t)$, in  
\cite{Haber:1996fp} it was shown 
that the leading (EFT) two--loop contributions are essentially absorbed by the one--loop
expression \cref{eq:AppRC}, if setting 
$\overline{m}_t(Q_t\equiv (\overline{m}_t M_S)^{1/2})$
and $\overline{m}_t(Q_t\equiv M_S)$,
respectively for the
term $ \sim \ln
M^2_S/\overline{m}_t^2$ 
and mixing terms $\sim \overline m_t X_t$.
Given that we do not seek the best possible approximation, the choice of a (unique) EWSB scale $Q_{EWSB}\simeq (m_{\tilde{t}_1} m_{\tilde{t}_2})^{1/2}\sim M_S$
for all running parameters
(or 
any fixed scale sufficiently close to the latter, as often conveniently chosen  
in Suspect and similar 
codes\cite{Skands:2003cj})
appears to be a
reasonable compromise. 



\item \cref{eq:AppRC} can also be derived from the diagrammatically obtained $\Pi_{ij}(0)$ self-energies and $t_i$ tadpoles, provided one carefully identifies\cite{Carena:2000dp} approximations at the same level
as described above.
\end{itemize}
There exists several refinements of \cref{eq:AppRC},
e.g. including sbottom~\cite{Haber:1996fp}, stau and QCD leading effects~\cite{Heinemeyer:2000nz}. In the following we stick to \cref{eq:AppRC}, referring to it as {\sl approximate} one--loop whose
simplicity is important for our construction, as explained next.


In order to determine $A_t$ as a function of the ``physical''
mass $m_h$, \cref{eq:AppRC} is solved as being 
a quadratic equation in $X^2_t$.

\begin{figure}[htb]
    \centering
    \includegraphics[width=0.6\textwidth]{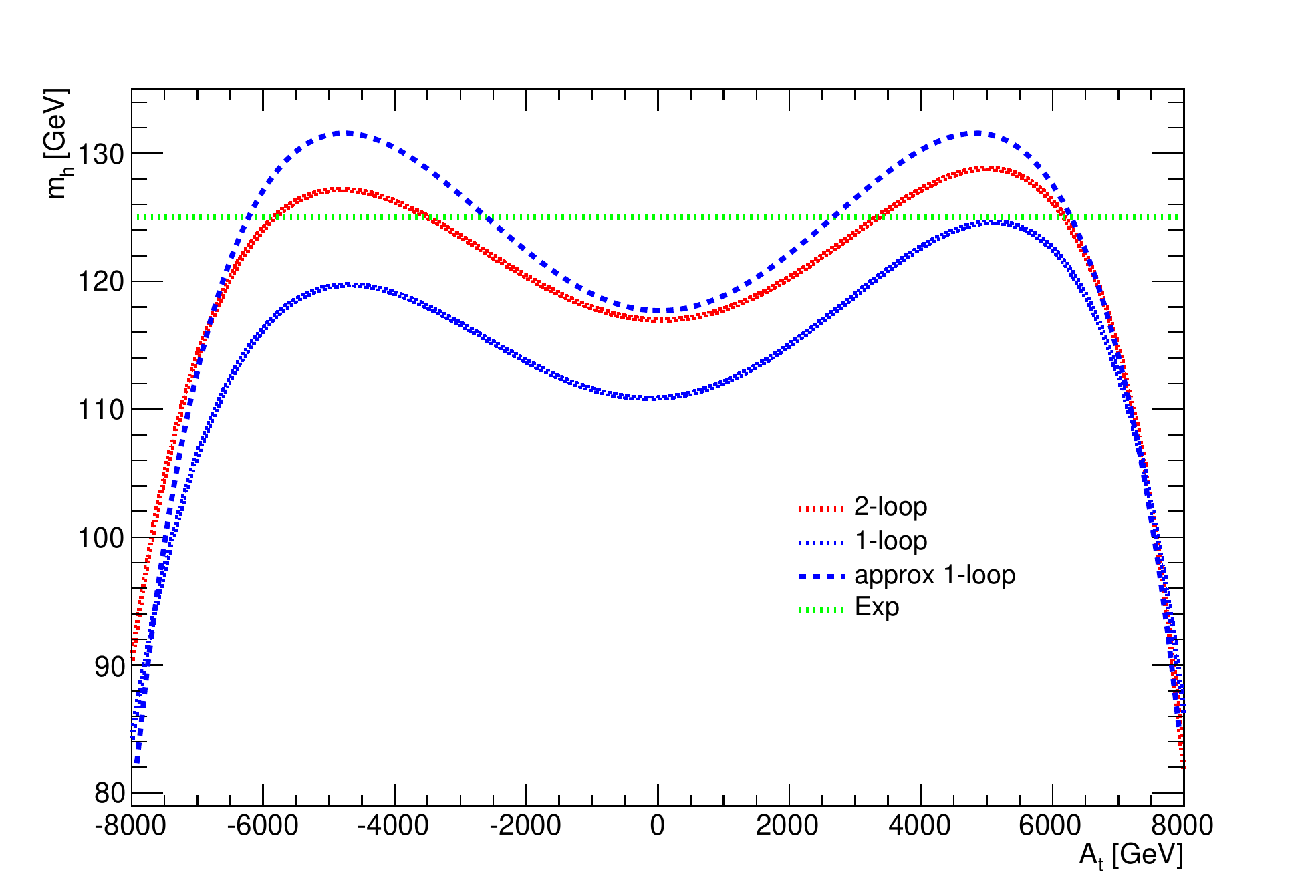}
    \caption{The prediction of $m_h$ at one-- and two--loop precision as well as for the approximate one--loop is shown as function of $A_t$ together with the measured mass.}
    \label{fig:mHAt}
\end{figure}

The procedure captures an essential feature of
the inversion for $A_t$: in principle there are two $X^2_t$ solutions which, when real-valued and positive, lead to
up to four different $A_t$ solutions. 
This well-known feature is illustrated in \cref{fig:mHAt}
where we show the $m_h$ prediction for the full one--loop, two--loop and the approximate one--loop calculations. For the 
$m_h(A_t)$ full one--loop and two--loop results, $A_t$ was 
varied and for each $A_t$ the EWSB minimisation was performed to derive the running
mass of the CP--odd Higgs boson and the $\mu$ parameter. All other parameters were fixed 
to their cliff point values. For the approximate one--loop, the coefficients of \cref{eq:AppRC} were calculated for the cliff point, thus fixing $\overline{m}_A^2$ and $\mu$ and varying only $A_t$ in the square root of \cref{eq:AppRC}
 shown in the figure. For the measured Higgs boson there are four intersections with the prediction. The third intersection (from left to right) with the two-loop prediction is consistent with the $A_t$ cliff point value of 3.61~TeV.

\begin{table}[htb]
\begin{center}
   \caption{The result of the $A_{t}$ determination from $m_{h}$ using \cref{eq:AppRC} is shown. s3 corresponds to the nominal solution: the cliff point.}\label{table:AppAtComp}
  \begin{tabular}{ccc}
  \toprule 
  s1 & $A_t$ [TeV] & -5.44 \\
  s2 & $A_t$ [TeV] & -3.61 \\
  s3 & $A_t$ [TeV] &  2.87 \\
  s4 & $A_t$ [TeV] &  6.36 \\
  \botrule
  \end{tabular}
    \end{center}
\end{table}
 
We turn now to the inverse approach of determining $A_t$ from  \cref{eq:AppRC} when the measured $m_h$ is taken as input.
The $A_{t}$ determination is illustrated in \cref{table:AppAtComp}. 
Four solutions are determined for $A_t$
as expected. The third solution s3 corresponds to the nominal solution: the cliff point. In each one of the solutions $\overline{m}_A^2$ and $\mu$  are those corresponding to the intersection of the two---loop Higgs mass with the measured $m_h$. 

For the nominal $A_t$ in s3 \cref{eq:AppRC} leads to a Higgs mass larger than the two--loop Higgs mass prediction (see Fig.\ref{fig:mHAt}). This is compensated
by an $A_t$ value smaller than the nominal one: 
2.87~TeV instead of 3.61~TeV.
The use of the measured $m_h$ in the calculation therefore inevitably leads to a shift in $A_t$. Its calculated value is within 25\% of the nominal value, a clear improvement compared to a blind guess.
At the same time, it clearly illustrates that an accurate $A_t(m_h)$ determination requires a more elaborate 
construction as compared to \cref{eq:AppRC}, which we will describe next.

\subsection{Full One--Loop: the HiggsMolar Function \label{subsec:FullOneLoop}}

\begin{figure}[htb]
\begin{center}
\includegraphics[width=0.45\textwidth]{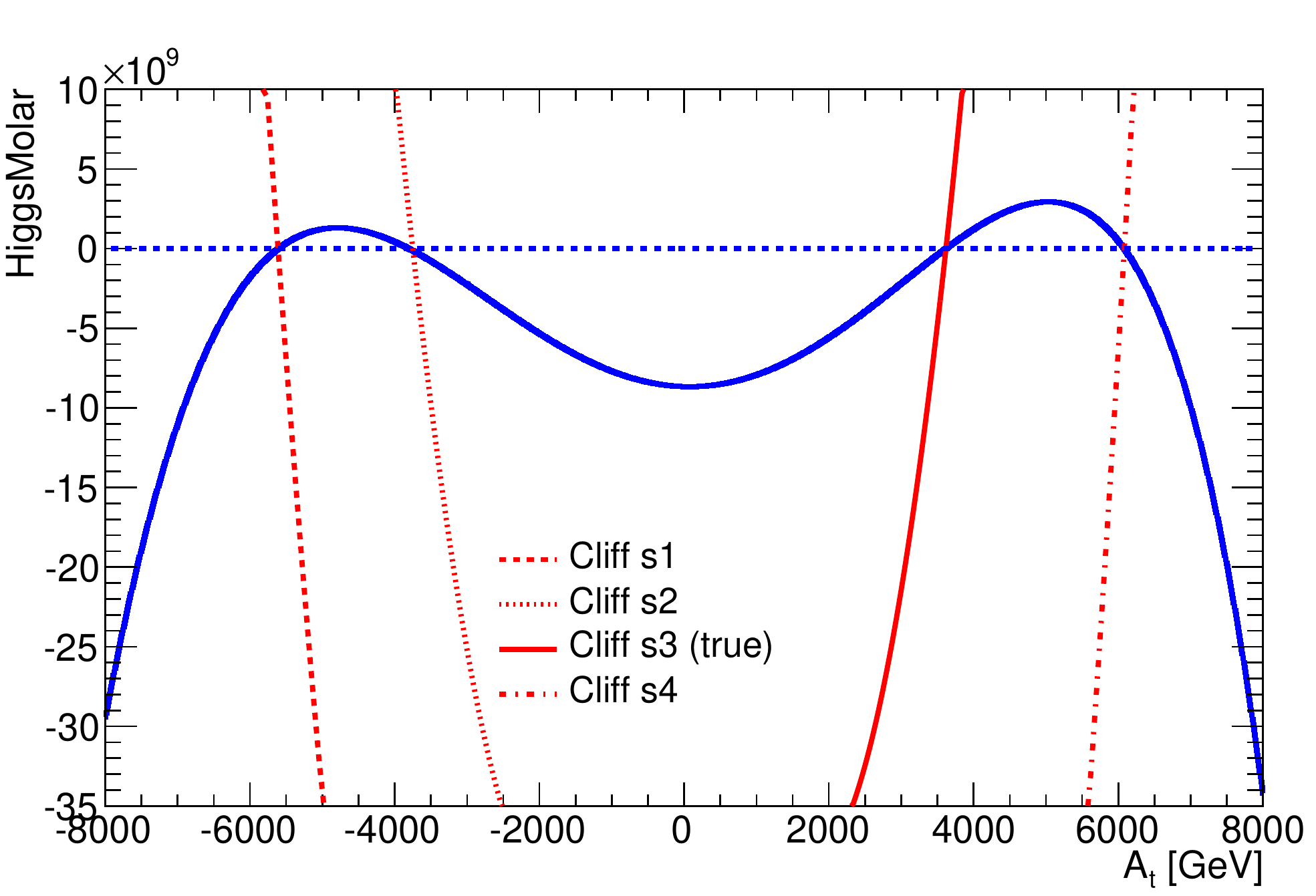}
\end{center}
\caption{The distribution of the \higgsmolar{} is shown as function of $A_t$: calculating the coefficients of \cref{eq:mh4At} for each $A_t$ (solid blue curve); 
fixing the coefficients of \cref{eq:mh4At} at the four $A_t$
solutions (red curves). For the four solutions, given by the intersections with the dotted blue line, only the $A_t$ range around the solution is shown by the red curves. 
}
\label{fig:higgsmolar}
\end{figure}

For the full one--loop inversion the starting point is the (formally exact) \cref{eq:mhpole} which we rewrite here as an essentially quadratic equation in the CP-even Higgs squared masses,

\begin{equation}
\begin{aligned}
 m_{h,H}^{4}- & m_{h,H}^{2}\left(({\cal M}^2_s)_{{}_{11}}(p^2)+({\cal M}^2_s)_{{}_{22}}(p^2)\right)  \\
 &+({\cal M}^2_s)_{{}_{11}}(p^2) ({\cal M}^2_s)_{{}_{22}}(p^2) - \left(({\cal M}^2_s)_{{}_{12}}(p^2)\right)^{2} \vert_{p^2 = m^2_{h,H}}=0 , \label{eq:mh21}
 \end{aligned}
 \end{equation}
 of which $m_h^2$ corresponds to the lighter mass solution. Here $({\cal M}^2_s)_{{}_{ij}}(p^2)$ denotes the {\sl real} parts of the matrix element, and the solutions $p^2=m_{h,H}^2$ are real-valued.\footnote{\label{footnote:real}Since the matrix 
 elements of $({\cal M}^2_s)(p^2)$ develop imaginary parts, 
 the squared pole masses are, strictly speaking, given by the real parts of the two solutions of  \cref{eq:mhpole}. In practice, though, neglecting the imaginary parts in the equation itself is a very good approximation for $m_h$, since the induced relative error on its estimate (of order $\frac{\Gamma_h}{m_h}$ with $\Gamma_h$ the total width), is negligibly small compared to other (higher order) theoretical uncertainties.}
 The $({\cal M}^2_s)_{{}_{ij}}(p^2)$'s extracted from \cref{eq:Ms} include perturbatively complete  loop corrections 
 where the full one--loop Higgs boson self-energies, $\Pi_{ij}$, and tadpoles $t_{i}$  are taken from 
 \cite{Pierce:1996zz}. In the $\dr$ scheme these expressions include contributions from all (s)particles (running) masses, 
resulting in a highly nonlinear dependence on $A_{t}$ from the stop sector. In particular 
the finite part of the 
one--loop scalar function $A_{0}$, occurring in the tadpoles and self-energies, has a rather involved $A_{t}$ dependence when its argument is 
the stop mass:
\begin{equation}
\label{eq:pisisjAt1}
\begin{aligned}
A_{0}(m_{\tilde{t}_{i}})&= m_{\tilde{t}_{i}}^{2}\left(1-\ln\left( \frac{m_{\tilde{t}_{i}}^{2}}{Q^{2}}\right) \right)
\end{aligned}
\end{equation} 
where $Q$ is the $\dr$-scheme 
renormalization scale, and where from the stop sector diagonalization, 
the $\dr$-scheme running stop masses can be written as
\begin{align}
\label{eq:mstops}
&m_{\tilde{t}_{1,2}}^2 = \frac12 \left( M^2 \mp \sqrt{a_s A_{t}^{2}+b_s A_{t} +c_s} \right) , 
\end{align}  
where
\begin{equation}
\begin{aligned}
\label{eq:asetc}
&M^2 = m_{\tilde{q}3_L}^{2} + m_{\tilde{t}_{R}}^{2} + 2 m_t^2 + \frac{1}{2} m_Z^2 \cos{2\beta}, \\
&a_s = 4 m_t^2, \\
&b_s = -8  m_t^2 \mu \cot\beta, \\
&c_s = \left(m_{\tilde{q}3_L}^{2} - m_{\tilde{t}_{R}}^{2} + (\frac{1}{2}-\frac{4}{3} s_W^2 ) m_Z^2 \cos{2\beta}\right)^2 + 4  m_t^2 \mu^2 \cot^2\beta , 
\end{aligned}
\end{equation}
are $A_t$ independent.\footnote{Note that $M^2$ gives the accurate
combination entering the exact one--loop expressions considered here.  $M_S^2$ in \cref{eq:MSusy} is in general obviously unequal to $\frac12 M^2$, unless 
$m_{\tilde{q}3_L}^2 = m_{\tilde{t}_{R}}^{2}$
 and the D-terms are neglected.
We use it in \cref{eq:AppRC} as a practical
approximation for $m_h$.}
The strategy is to rewrite \cref{eq:mh21} as an equation for $A_t$ by 
extracting from \cref{eq:Ms} the explicit polynomial $A_{t}$ 
dependencies or "power counting" within each $\Pi_{ij}$ and $t_{i}$. 
On close inspection of the various contributions in e.g. \cite{Pierce:1996zz} we identify:
\begin{enumerate}
 \item terms depending linearly on $A_{t}$ (with coefficients identified by a superscript '1'), 
 originating from the $s_k$--$\t t_i$--$\t t_j$ couplings, 
 $g_{s_2\t t_1\t t_1}$, $g_{s_2\t t_2\t t_2}$,
 $g_{s_2\t t_1\t t_2}$ which are given by
\begin{equation}
\label{eq:gs2}
\begin{aligned}
g_{s_2\t t_1\t t_1} &= c_t^{2}  \, g_{s_2{\tilde{t}_L}{\tilde{t}_L}} + 2 c_t s_t \, g_{s_2{\tilde{t}_L}{\tilde{t}_R}} + s_t^{2} \, g_{s_2{\tilde{t}_R}{\tilde{t}_R}},\\
g_{s_2\t t_2\t t_2} &= s_t^{2} \, g_{s_2{\tilde{t}_L}{\tilde{t}_L}} - 2c_t s_t \, g_{s_2{\tilde{t}_L}{\tilde{t}_R}} + c_t^{2} \, g_{s_2{\tilde{t}_R}{\tilde{t}_R}}, \\
g_{s_2\t t_1\t t_2} &= s_t c_t \, (g_{s_2{\tilde{t}_R}{\tilde{t}_R}} - g_{s_2{\tilde{t}_L}{\tilde{t}_L}}) + (c_t^{2}-s_t^{2}) \, g_{s_2{\tilde{t}_L}{\tilde{t}_R}},\\
g_{s_2{\tilde{t}_L}{\tilde{t}_R}} &= \frac{y_{t}}{\sqrt{2}}A_{t},
   \end{aligned}
\end{equation} 
with $y_t$ the top quark Yukawa coupling, $s_t\equiv \sin\t \theta_t$, $c_t\equiv \cos\t \theta_t$, where $\t \theta_t$ denotes the stop mixing angle, 
$\t t_1, \t t_2$ are the stop mass eigenstates, 
$\t t_L$, $\t t_R$ the gauge eigenstates, and $s_k$ are the neutral scalar states in the basis corresponding to \cref{eq:Ms}. Note that
$g_{s_2{\tilde{t}_R}{\tilde{t}_R}}$, 
$g_{s_2{\tilde{t}_L}{\tilde{t}_L}}$ do not depend
explicitly on $A_t$, their expressions can be found
e.g. in \cite{Pierce:1996zz} (denoted by $\lambda_{s_2{\tilde{u}_{L,R}}{\tilde{u}_{L,R}}}, \lambda_{s_2{\tilde{u}_L}{\tilde{u}_R}}$ therein).
\item terms depending quadratically  on $A_{t}$ (with coefficients identified by a superscript '2'), 
originating from the $g_{s_2t_it_i}^2$ that appear
solely in $\Pi_{22}(p^2)$;
\item terms depending on $\sqrt{a_s A_{t}^{2}+b_s A_{t} +c_s}$  (with coefficients identified by 's')
originating from the $m_{\tilde{t}_{i}}^{2}$ in $A_{0}(m_{\tilde{t}_{i}})$;
\item  a term  containing $A_{t}\cdot\sqrt{a_s A_{t}^{2}+b_s A_{t} +c_s}$ (with coefficient identified by '1s')
resulting from the occurrence of the product $g_{s_2t_it_i}\times A_{0}(m_{\tilde{t}_{i}})$ appearing solely in $t_2$;
\item finally all remnant contributions with no (explicit) dependence or with logarithmic dependence on $A_{t}$, are identified by a '0' superscript.
It is an important part of our strategy that in our $A_t$-power counting any ``logarithmic" dependence $\sim 
\ln (M^2 \mp \sqrt{a_s A_t^2+\cdots})$
on $A_t$ (such as in the second term of \cref{eq:pisisjAt1}, and in the one--loop scalar function $B_0$), as well as the other algebraic dependence 
on $A_t$ in $B_0$, are
incorporated exactly as they stand  
within the coefficients of the above listed relevant $A_t$ powers. In the following we dub these dependencies ``residual''. The dependence of the stop mixing angle $\t \theta_t$ on $A_t$  is also treated as residual, since it enters in \cref{eq:gs2} through $c_t$ and $s_t$  that remain obviously bounded functions of $A_t$.
\end{enumerate}
According to the previous $A_t$ power counting, within the relevant (one--loop) $\Pi_{ij}$
and $t_i$ individual contributions, there are no higher
degree monomials in $A_t$ than the $A_t^k$ identified above
with $k \le 2$.
This gives the following formal decomposition of the tadpoles and self-energies:
\begin{align}
&\frac{t_{1}}{v_1}= t_{1}^{ (s) }\sqrt{a_s A_{t}^{2}+b_s A_{t} +c_s}+ t_{1}^{ (0) } \nonumber\\
&\frac{t_{2}}{v_2}= t_{2}^{ (1s) } A_t \sqrt{a_s A_{t}^{2}+b_s A_{t} +c_s}+ t_{2}^{ (1) } A_t + 
t_{2}^{ (s) }  \sqrt{a_s A_{t}^{2}+b_s A_{t} +c_s}  + t_{2}^{ (0) }\nonumber\\
&\Pi_{11}= \pi_{11}^{(s)} \sqrt{a_s A_{t}^{2}+b_s A_{t} +c_s}+ \pi_{11}^{ (0) }   \label{eq:pisisjAt}\\
&\Pi_{12}= \pi_{12}^{ (1) } A_t + \pi_{12}^{ (0) }\nonumber\\
&\Pi_{22}= \pi_{22}^{ (2) } A_t^2+ \pi_{22}^{ (1) } A_t+    \pi_{22}^{ (s) } \sqrt{a_s A_{t}^{2}+b_s A_{t} +c_s} +\pi_{22}^{ (0) } \nonumber
\end{align}
\Cref{eq:pisisjAt} is simply a convenient 
rewriting of already available exact one--loop expressions, no contributions are ignored.
Using \cref{eq:pisisjAt} to display the algebraic dependence on $A_t$ in \cref{eq:mh21}, the following molar-shaped
function $\higgsmolar(A_{t})$, which should consistently vanish for any $A_{t}$ solution, is obtained:
\begin{align}
 \higgsmolar(A_{t})= & C_{3} A_{t}^{3} + C_{2} A_{t}^{2} + C_{1} A_{t} + C_{0} \nonumber \\ 
   &+ (R_{2} A_{t}^{2} + R_{1} A_{t}+ R_{0}) \sqrt{a_s A_{t}^{2} +b_s A_{t}+ c_s}=0 \label{eq:mh4At}.
\end{align}
The \higgsmolar{} is shown as function of $A_t$ in \cref{fig:higgsmolar}. For each $A_t$ the coefficients are
recalculated. The four other curves show the behavior of the function if the coefficients determined at the intersections
of the \higgsmolar{} with the line $\higgsmolar = 0$ are used in the vicinity of the intersections. The curves illustrate the variation of the coefficients. 

This equation can in principle be used separately either for the lighter or for the heavier
CP-even Higgs masses, as clear from \cref{eq:mh21}. Hereafter we are only considering the lighter Higgs mass at $\sim 125\mathrm{GeV}$ as input. The $C_i$ and $R_i$ coefficients are easily identified upon use of \cref{eq:pisisjAt,eq:Ms,eq:mh21}. They contain all the residual $A_t$ dependencies quoted above, neglecting all imaginary parts as justified in \cref{footnote:real}.

The cubic term results  from the product $(t_1-\Pi_{11}) \times t_{2} $ present in  $({\cal M}^2_s)_{11} ({\cal M}^2_s)_{22}$, with a coefficient $C_3$ given by
\begin{equation}
\label{eq:coef}
C_{3}=  4\overline m_t^2 ( t_{1}^{(s)}  - \pi_{11}^{(s)}  ) t_{2}^{(1s)}.
\end{equation}

In particular, 
in contrast with \cref{eq:AppRC} and unless performing in \cref{eq:mh4At} an expansion in $X_t$,
explicit $X_t^4\sim A_t^4$ terms cannot occur within the exact \cref{eq:mh21}
as the equation does not involve squares of terms of the type '2' or '1s'.

The general form of the other $C_{i}$ and $R_{i}$ coefficients is given in \cref{app:higgsmolar}. These coefficients have a more involved dependence on (differences of) the quantities $\pi_{ij}^{(0,s,1,2)}$ and $t_{i}^{(0,s,1,1s)}$ entering \cref{eq:pisisjAt}. The relevant one--loop expressions of the latter are also given in \cref{app:higgsmolar}. They allow to track the residual dependence on $A_t$ and the absence of some finite combinations in relation to the expected cancellation of the quadratic divergences in softly-broken SUSY. 
The full one--loop
explicit dependence on $A_t$ is thus included in \cref{eq:mh4At}. A further implicit dependence on $A_t$ will originate from the two EWSB conditions  Eqs.~(\ref{eq:MAEWSB}),~(\ref{eq:muEWSB}) when imposed beyond the tree-level, due to the presence of $\overline m_A^2$ as well as $\mu$ in the $C_{i}$ and $R_{i}$ coefficients. In particular, the $t_2$-tadpole dependence in $\overline m_A^2$
will induce, through the term $\overline{m}_{11}^2 \times t_2$ appearing in the cross-product $({\cal M}^2_s)_{11} ({\cal M}^2_s)_{22}$ in \cref{eq:mh21}, an effectively {\sl quartic} dependence on $A_t$ for large $A_t$, not
explicit in \cref{eq:mh4At}. 
This entails
solving simultaneously \cref{eq:mh4At} and the EWSB constraints Eqs.~(\ref{eq:MAEWSB}) and (\ref{eq:muEWSB}), which we will perform numerically in a consistent way as described in 
\cref{sec:algo}. Other implicit dependencies on $A_t$ are discussed in \cref{subsec:HigherLoops}.
Hereafter we ignore momentarily these issues and focus solely on the resolution of \cref{eq:mh4At}.  

  

\subsection{Full One-Loop Inversion: the Fixed Point Algorithm \label{subsec:FullOneLoopFP}}

To solve {\cref{eq:mh4At} for $A_t$, a fixed point iterative method is used.
 For this purpose we define 
 \begin{equation}
\begin{aligned}
\cubicFP(A_t)= -\frac{1}{C_{3}} \Big( & C_{2} A_{t}^{2} + C_{1} A_{t} + C_{0} \\
& + (R_{2} A_{t}^{2} + R_{1} A_{t}+ R_{0})\sqrt{a_s A_{t}^{2} +b_s A_{t}+ c_s} \Big), \\
\end{aligned}     
 \end{equation}
 and rewrite \cref{eq:mh4At} as
 \begin{equation}
 A_t =  \sqrt[3]{\cubicFP(A_t)}.
 \end{equation}
 It is then clear that finding all the real-valued solutions  of \cref{eq:mh4At} is equivalent to determining all the fixed points $A_t=A_{t}^{FP}$, 
 satisfying $\linearFP(A_{t}^{FP})= A_{t}^{FP}$, of the function \linearFP{} defined by
 \begin{equation}
\linearFP(A_t) \equiv  \sqrt[3]{\cubicFP(A_t)},
\label{eq:linearFP}
 \end{equation}
 with only {\sl real-valued} cubic roots allowed.


\begin{figure}[htb]
\begin{center}
\includegraphics[width=0.45\textwidth]{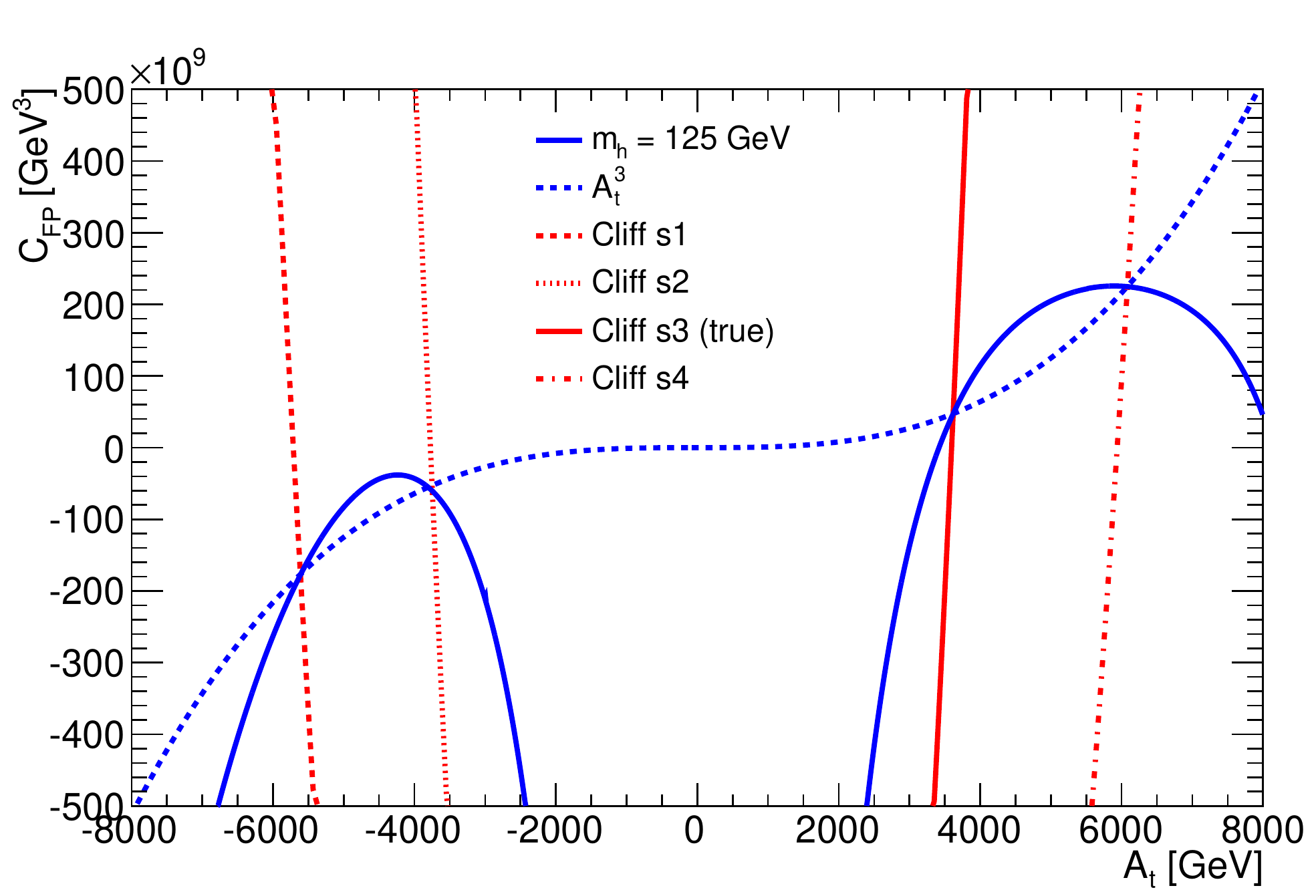}
\includegraphics[width=0.45\textwidth]{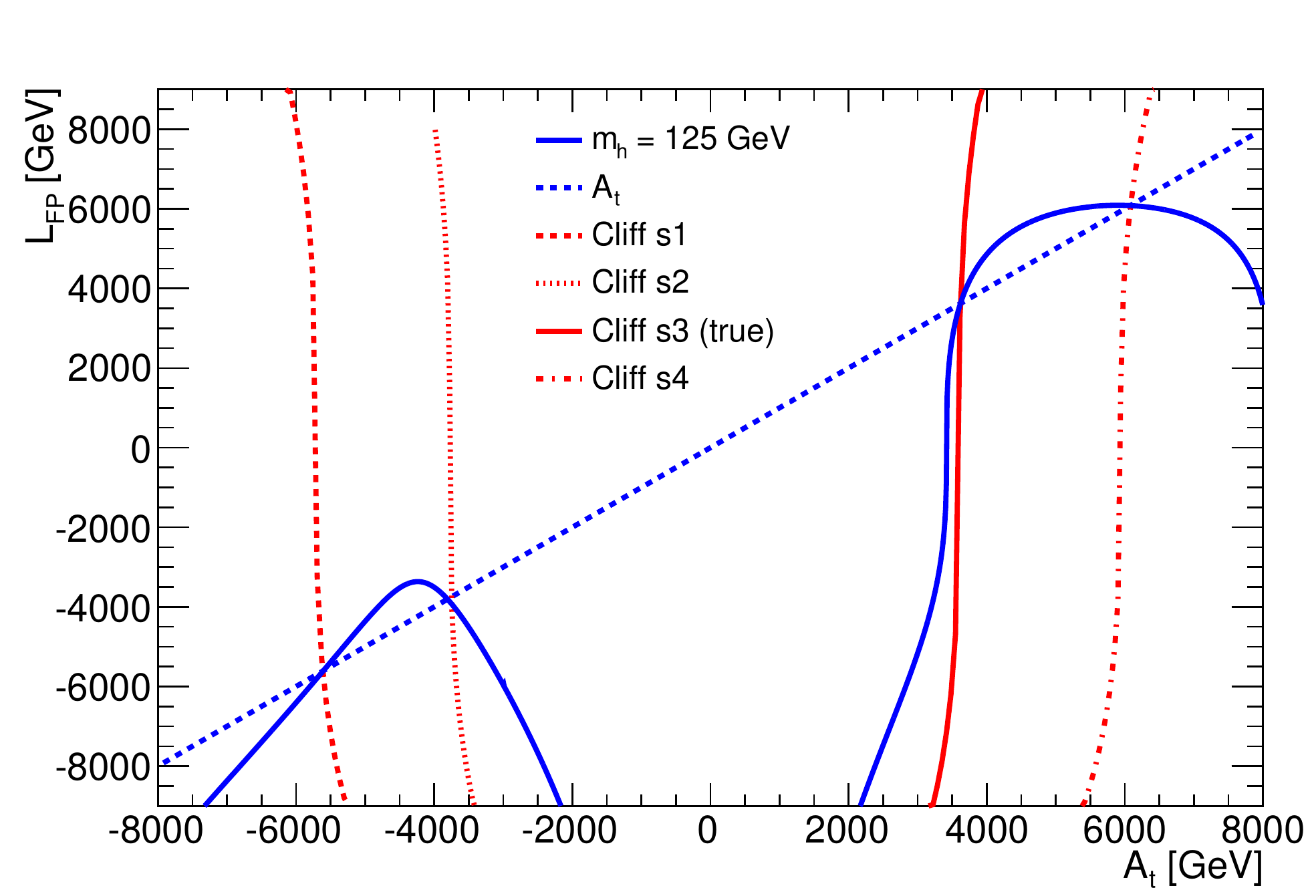}
\end{center}
\caption{The distributions of the functions \cubicFP{} and \linearFP{} are shown as function of $A_t$: calculating the coefficients of \cref{eq:mh4At} for each $A_t$ 
(solid blue curve); 
fixing the coefficients of \cref{eq:mh4At} at the four $A_t$
solutions (red curves). For the four solutions, given by the intersections with the dotted blue line, only the $A_t$ range around the solution is shown by the red curves. 
}
\label{fig:FP}
\end{figure} 
To determine the fixed points 
one starts from a guess value
 $A_{t, 0} = A_{t}^{Guess}$ and studies the convergence of the sequence $A_{t, i+1} = \linearFP(A_{t_{i}})$. The iterations return unambiguously real-valued $A_t$ as a consequence of the definition of \linearFP, \cref{eq:linearFP}. As we will specify in more detail in \cref{sec:Algo}, appropriate $A_t$-guess values, not too far from the exact solutions, are those obtained from our approximate one-loop \cref{eq:AppRC} that already captures the multi-solution structure. However, even if starting relatively close to the exact solutions, the method will catch only the {\sl attractive} fixed points and can thus miss some, otherwise acceptable, $A_t$
solutions. }
 

 One expects typically four distinct solutions as illustrated in \Cref{fig:FP} for the stop cliff benchmark: The functions $\cubicFP(A_t)$ and $\linearFP(A_t)$, in blue, intersect  
the dotted blue lines at the fixed points of these functions, corresponding to the four $A_t$ solutions that are
consistent with $m_h=125\mathrm{GeV}$. For the blue curve the coefficients $C_i$ and $R_i$ of the two fixed point functions were
recalculated at each $A_t$. For the four red curves these coefficients are frozen at their values
calculated at the intersection of the blue curve with the blue dotted line, for $A_t$ varying in
the vicinity of the intersection.

The dotted blue line on
\Cref{fig:FP}-right being the bisector, 
it follows that the slope of $\linearFP$ at its fixed points is alternatively greater or smaller 
than one. We label the solutions $s1, s2,s3,s4$ ordered in ascending in $A_t$. Given that $\linearFP$ takes negative values for very large 
$\lvert A_t \rvert$, the slopes of $\linearFP$ at $s1$ and $s3$ are necessarily greater than one and those at $s2$ and $s4$ less than one. $s1$ and $s3$ are thus {\sl repulsive} fixed 
points. In contrast, $s2$ and $s4$ will be either {\sl attractive} if the slope is between
$-1$ and $1$, or  {\sl alternating/repulsive} if the slope is less than $-1$. The slope at $s2$ is typically less than zero due to the strong variation of $\linearFP$ in the vicinity of
$\lvert A_t \rvert\sim 0$.
It follows that the iterative procedure on $\linearFP$ as described above will always miss solutions $s1$ and $s3$  while capturing at best
 two solutions, namely $s2$ and $s4$, provided furthermore the initial guess values $A_{t}^{Guess}$ close enough to
 $s2$ and $s4$ to feel their attracting character. If a guess point is chosen anywhere above $s3$
 the iterations will repel it from $s3$ and converge to $s4$. If chosen  below $s1$ it will be repelled to the left and the
 procedure will never converge. Finally, if chosen between $s1$ and $s3$ then, depending on the behaviour around $s2$, the convergence to $s4$ (and occasionally to $s2$ if its slope is greater than $-1$) may or may not occur.
 
 In order to ensure convergence and be able to capture all solutions, we consider
 a generalized family $\linearFP_{\tau}$ of functions parametrized by $\tau \neq 0$ as follows:
 \noindent
\begin{equation}
\label{eq:t}
\linearFP_\tau(A_t) = \frac{1}{\tau} (\linearFP(A_t) -A_t) + A_t.
 \end{equation}
 
 Any fixed point of \linearFP{} is also a fixed point of $\linearFP_\tau$ and vice versa, for any value of $\tau$.
 
 Moreover from 
 \begin{equation}
\label{eq:dt}
 \linearFP'_{\tau}(A_{t})= 1   + \frac{\linearFP'(A_{t}) -1}{\tau},
 \end{equation}
   follows that one can always choose $\tau$ in such a way that $\linearFP'_{\tau}(A_{t}^{FP})$ lies in the interval $[-1,1]$
   for any given value of $\linearFP'(A_{t}^{FP})$.
  Thus the advantage of $\linearFP_\tau$ is that all fixed points of $\linearFP$ can be made {\sl attractive} with respect to $\linearFP_\tau$ for appropriately chosen values
of $\tau$.
 In this case an iteration over the sequence
 $A_{t, i+1} = \linearFP_{\tau} (A_{t_{i}})$ is guaranteed to converge on $A_{t}^{FP}$, at least when the initial guess values
are not too far from the solution. However, $A_{t}^{FP}$
 is not known in advance, even less $\linearFP'(A_{t}^{FP})$. Without this knowledge, a rough strategy 
 to converge on a given solution could be to choose:  
\begin{itemize}
    \item[-] $\tau \gtrsim 1$ for $s4$ 
    \item[-] $\tau \ll -1$ for $s1, s3$
    \item[-] $\tau \gg 1$ for $s2$.
\end{itemize}
Actually one can do better by determining optimal values of $\tau$ from the knowledge of the local 
variation of $\linearFP$ during the iterative procedure. 
A numerical estimate $\Delta \linearFP/\Delta A_t$ of the first derivative of $\linearFP$ with respect to $A_t$ can in principle be calculated at any given step during the iterations on $A_t$ at moderate computational cost. Then choosing the $\tau$ parameter as follows
\begin{align}
&\frac{\Delta \linearFP}{\Delta A_t} - 1 <  0 :  \tau = + \left\lvert\frac{\Delta \linearFP}{\Delta A_t} - 1 \right\rvert, \nonumber\\
&\frac{\Delta \linearFP}{\Delta A_t} - 1 > 0 :   \tau= - \left\lvert\frac{\Delta \linearFP}{\Delta A_t} - 1 \right\rvert,
\label{eq:algotau}
\end{align}
ensures that an estimate of the derivative of $\linearFP_\tau$ is close to zero, cf. \cref{eq:dt}. Since this estimate is in practice not at the fixed point,  \cref{eq:algotau} is not sufficient to guarantee convergence. However, it approximates a necessary condition, fulfilling a convergence criterion discussed in \cref{app:convergence}, see \cref{eq:conv-cond}. This allows to make any fixed point attractive, provided initial guess values are not too far from that fixed point.
A simple and efficient algorithm will be implemented along these lines, as described in \Cref{sec:Algo}.


\subsection{Dominant Two--Loop Inversion\label{subsec:FullTwoLoop}}

The extension of the above exact one--loop method to the two--loop contributions is 
rather straightforward.
The latter corrections depend dominantly on the strong, weak  and third generation Yukawa couplings $y_i$, leading
to terms of $\mathcal{O}$($ \alpha_{s,i}$)$\times \mathcal{O}$(One--loop),
where $\alpha_i\equiv y_i^2/(4\pi)$. In the standard fixed-order (diagrammatic) $m_h$
calculations,
$\Pi_{ij}^{(2-loop)}$ and $t_i^{(2-loop)}$ contributions are  added to  the ${\cal M}^2_s$ matrix elements
which enter the squared Higgs boson mass equation  shown 
in \cref{eq:mh21}.

Due to the extra loop suppression factor, 
at any scale relevant for the MSSM spectrum calculation
where either $\alpha_s(Q)$ or $\alpha_i(Q)$ remain moderate, 
the two--loop contributions are a moderate correction relative to  
the one--loop contributions (even if including those is obviously 
very relevant for a more precise comparison between the MSSM prediction and the 
measured $m_h$ value. For instance typically for the 
benchmark cliff point, restricting to exact one--loop would give 
$m_h\sim 118$ GeV instead of $\sim 125$ GeV as in Table \ref{tab:MhInvPoints}.)
Thus, for the $A_t(m_h)$ inversion, rather than trying to extract specific quite involved $A_{t}$ 
dependencies from the $\Pi_{ij}^{(2-loop)}$ and $t_i^{(2-loop)}$ contributions, the latter are incorporated just as they contribute to Eq.(\ref{eq:Ms}): more precisely, 
within the above algorithm the available two--loop contributions are formally treated
as if they were independent of $A_{t}$, therefore concretely incorporated as additional contributions to either $t_i^{(0)}$ or  $ \pi_{ij}^{(0)}$ 
in \cref{eq:pisisjAt}:

\begin{equation}
\label{eq:2loop}
t_i^{(0)} \to t_i^{(0)} + t_i^{(2-loop)},\;\;
\pi_{ij}^{(0)} \to \pi_{ij}^{(0)} + \Pi_{ij}^{(2-loop)}.
\end{equation}

This is then corrected iteratively for the true $A_t$ dependence. 
The seven coefficients entering the \higgsmolar($A_{t}$) 
function of \cref{eq:mh4At} can now be computed incorporating 
consistently two--loop contributions.

 \subsection{Including Higher Orders and Refinements}
 \label{subsec:HigherLoops}

At this stage the inversion is formally 'exact' at the considered
(perturbative) level
of theoretical precision taken for the $t_i$ tadpoles and self-energy contributions $\Pi_{ij}$, 
namely full one--loop and only the dominant two--loop contributions. 
It is a straightforward matter to incorporate either more complete
two--loop and/or higher (3--loop) contributions,
by considering those contributions similarly $A_t$ independent, since
their actual $A_t$ dependence, independent of its complexity, is screened by tiny 
perturbative expansion coefficients. 
As long as higher order corrections are obtained diagrammatically in the form of self-energy or tadpole contributions, these could be included explicitly by adding them
to the $t^{(0)}_i$ and $\pi^{(0)}_{ij}$ contributions.

It is well known that sizeable theoretical uncertainties in $m_h$ determinations (customarily taken as $\sim \pm 2-3$ GeV in phenomenological analyses) are due to presently unknown higher order contributions, discrepancies between different renormalization schemes, etc (see e.g. \cite{Degrassi:2002fi,Allanach:2004rh},
or for more recent analyses \cite{Allanach:2018fif,Bahl:2019hmm}, as well as the recent updated discussion in \cite{Slavich:2020zjv}). 
Given these uncertainties, one might question the importance of devising a very accurate inversion procedure. The answer is obvious: not to introduce artificially further uncertainties in the determination of $A_t(m_h)$ than there are from a given content
of higher order contributions included in the standard $m_h$ determination.
Related to this, there remains one subtlety to consider: Even at one--loop level, there
are extra implicit $A_t$ dependencies that would not be accounted for by the previous
algorithm, if one relied solely on the procedure leading to \cref{eq:mh4At}. Indeed,
the self-energies and tadpoles also depend typically on SM-like gauge and Yukawa couplings, 
as well as other running $\dr$ parameters ($s_W$, $m_W$), which are affected by threshold corrections,
 depending themselves on the MSSM parameters, therefore depending on $A_t$ 
 in a highly nontrivial way in this case.
While all these threshold corrections give contributions that are formally of higher 
(at least two--loop) order {\em within} the loop self-energy and tadpole expressions, they
can induce a numerical inconsistency bias if not incorporated in the inversion, slightly shifting the resulting $A_t(m_h)$ with respect to its actual 'reference' value in a standard top-down calculation. 
This, as well as the other residual or implicit dependencies on $A_t$ already mentioned in \cref{subsec:FullOneLoop},  are, however, consistently taken into account in the full algorithm as we explain next.

\section{The Full Inversion Algorithm} \label{sec:Algo}

The algorithm has been implemented in
SuSpect3~\cite{Djouadi:2002ze,Brooijmans:2012yi}. SuSpect3 is a public spectrum calculator for multiple
supersymmetric models that includes, within \cref{eq:Ms} for $m_h$, the $\dr$ radiative
corrections at full one--loop and dominant two-loop orders (involving for the latter the QCD and third family Yukawa contributions, but at vanishing $p^2$).
Other MSSM spectrum calculators are, non-exhaustively, 
SOFTSUSY~\cite{Allanach:2001kg}, SPHENO ~\cite{Porod:2003um,Porod:2011nf,Staub:2017jnp}, FeynHiggs~\cite{Heinemeyer:1998yj,Heinemeyer:1998np,Heinemeyer:2000nz,Bahl:2018qog}, and FlexibleSUSY~\cite{Athron:2014yba}.
Note that on top of fixed-order calculations including some contributions 
beyond the above mentioned two-loop order,
some of these codes (FeynHiggs, SPHENO, FlexibleSUSY) also include 
resummations of large logarithms in an EFT approach for 
the Higgs mass calculations,
thus with an a priori increased precision for large soft-supersymmetry breaking mass scenarios. 

Before describing the full inversion algorithm, let us first
briefly recall the procedure to determine the spectrum in the MSSM. This involves solving the RGE to evolve the parameters between the EWSB scale ($Q_\mathrm{EWSB}$) and the scale given by the mass of the Z~boson  ($m_Z$), as well as solving the EWSB equations \cref{eq:MAEWSB}, \cref{eq:muEWSB} at $Q_\mathrm{EWSB}$. 
Radiative corrections to the supersymmetric particle and Higgs boson masses are calculated at $Q_\mathrm{EWSB}$. 
Supersymmetric radiative corrections to Standard Model parameters, the most important in the present study being the 
top yukawa coupling, are calculated at $m_Z$.

For EWSB three variants have been implemented. 
If $\overline{m}^2_{H_d}$, $\overline{m}^2_{H_u}$ and the sign of $\mu$ are given as input, the running mass squared $\overline{m}_A^2$ and the Higgs mass parameter $\mu$ are calculated. Alternatively $\mu$ and either the pole mass $m_A$ 
or the tree-level running mass squared $\overline{m}_A^2$ can be given to calculate $\overline{m}^2_{H_d}$, $\overline{m}^2_{H_u}$.

Both the RGE evolution and the EWSB calculations are implemented as iterations. The convergence 
is tested on the stability of $\overline{m}_A^2$ or $\overline{m}_{H_u}^2$ between successive iterations. The
choice depends on the input parameter set chosen.

We recall that for a precise calculation of the MSSM
spectrum in the standard top-down approach, it is essential that some of the relevant running parameters at a given scale, and
consequently the physical (pole) masses, are calculated iteratively, as these parameters are nontrivially modified by radiative corrections, which 
in turn depend on potentially all MSSM parameters.\footnote{ 
In particular for determining $\mu$ from Eq.(\ref{eq:muEWSB}) since
the right hand side depends itself implicitly on $\mu$ from the tadpole
contributions.} There is also an iteration between 
the EWSB and $m_Z$ scales, since important 
radiative corrections, depending themselves on the MSSM spectrum, 
are incorporated to extract the $\dr$-scheme gauge and 
Yukawa couplings upon matching their experimentally measured values. 
The convergence criterion
for the RGE iteration depends on the choice for the EWSB parameters.

\subsection{Algorithm}
 \label{sec:algo}

In the $m_h$MSSM the $A_t(m_h)$ determination algorithm extends the previously described calculation of 
the relevant EWSB parameters of the MSSM in SuSpect3. 
The determination of $A_t$ has been added to this already necessarily iterative structure as the determination of an additional parameter.
The 
algorithm to determine 
$A_t$ is independent of the parameter input choice. In the following the numerical 
examples are given for an input of $\overline{m}^2_{H_d}$, $\overline{m}^2_{H_u}$ and the sign of $\mu$.
The calculation starts with the RGE evolution from the Z~boson scale to a high scale. $A_t$ is initialized arbitrarily to a fixed value (10~GeV)
as the parameter will be determined after the
RGE evolution to $Q_{\mathrm{EWSB}}$.
The procedure is similar to the initialization of $\mu$.

First the MSSM EWSB calculations are performed, i.e., $\mu$ and $\overline{m}_A^2$ are determined, and then $A_t$ is determined. 
This procedure is repeated until convergence is reached according to the MSSM criteria, i.e., $\mu$ is stable 
and therefore $\overline{m}_A^2$ is stable as well.

At the first and second RGE iteration the 
approximate one--loop \cref{eq:AppRC} with the measured $m_h$ as input is solved to extract a new $A_t$ as explained
in \cref{subsec:AppOneLoop}. Taking into account the newly determined $A_t$, the RGE evolution to the Z~mass scale is performed. 
Radiative corrections are calculated, in particular to the top yukawa terms. The parameters are then RGE evolved to 
the high scale. The second RGE iteration therefore
starts at the high scale with the $A_t$ value derived from the approximate one-loop algorithm in the first iteration.

For the cliff point, \cref{table:AppAtComp} shows that the correct
$A_t$ is within 25\% of the calculated value. The use of \cref{eq:AppRC} is preferred over the fixed point algorithm
as the radiative corrections used depend on $y_t \cdot \mu$. This allows to stabilize quickly $y_t$ with an approximate $A_t$ at low computational cost.
Using the full radiative calculations at this stage would lead to longer iterations as the variations of both $\mu$ and $A_t$ take longer
to stabilize.

For the third and all following RGE iterations the full radiative calculations are used 
for EWSB, combined with the fixed point algorithm described in \cref{subsec:FullOneLoopFP,subsec:FullTwoLoop} to determine $A_t$. For each new $A_t$, obtained from $\linearFP_\tau$, \cref{eq:t}, the tree-level stop sector and the Higgs sector including radiative corrections are recalculated. This has the advantage that not only the leading terms of $\linearFP$, \cref{eq:linearFP}, are taken into account, but also both the residual and implicit $A_t$ dependencies in the coefficients $C_i$ and $R_i$, as explained previously. The recalculation of the stop sector and the Higgs sector for each iteration brings the function from the red curves closer to the blue (nominal) curves in \cref{fig:higgsmolar,fig:FP}.

The convergence depends on the $\tau$ parameter. When the full radiative corrections are used in the algorithm, an estimate $\Delta \linearFP/\Delta A_t$ of the first derivative of $\linearFP$ with respect to $A_t$ is calculated numerically at the end of the iteration on $A_t$. The iterations are stopped when the relative change of $A_t$ between the last and the current value is smaller than a threshold. Numerical values are given below.
The $\tau$ parameter is then adjusted according to \cref{eq:algotau} 
which is used in the next determination of $A_t$.


\subsection{Proof of Concept}

\begin{table}[htb]
\begin{center}
    \caption{The result of the Higgs inversion algorithm is shown for the stop cliff. The true point is s3. As a cross check the $m_h$ values shown are the result of the standard $m_h(A_t)$ calculation with the $A_t$ obtained from the input $m_h$ parameter.}\label{tab:HiggsInvResult}
    \begin{tabular}{lcccc}
         \toprule
         stop cliff &  s1 & s2 & s3 & s4 \\
         \midrule
         $A_t$ [GeV] & -5617.3   & -3796.1  & 3609.7   & 6082.5 \\
         $m_h$ [GeV] &   125.012 &  125.012 &  125.012 & 125.012 \\
         \botrule
    \end{tabular}
    \end{center}
\end{table}

As proof of concept the calculation is performed for all four possible solutions using the full algorithm. 
The $A_t$ is first determined via the approximate one-loop calculation for the first two RGE iterations 
and then the fixed point algorithm is used for all following iterations. $A_t$ is refined at each step.
To converge on the
spectrum calculation the RGE iterations are stopped once $\overline{m}_A$ is stabilized 
to the permil level. The EWSB iterations are stopped when $\mu$ has converged at permil level. 
The iterations on $A_t$ are run with a convergence criterion of permil.

The results of the full algorithm are shown in \cref{tab:HiggsInvResult}. Four solutions are obtained as expected. The calculation of $A_t$ and the calculated $m_h$ are in excellent agreement with the expected values.

In point s3 the deviation of the calculated $A_t$ value from the expected value in \cref{tab:MhInvPoints} is far 
smaller than convergence criterion on $A_t$ suggests. This is the consequence of the hierarchical structure of the iterations. The iteration
on $A_t$ is at the lowest level, therefore the calculation of $A_t$ is also refined for each EWSB and (times) RGE iteration until the running mass of the A~boson and the $\mu$ parameter have converged. This leads to a higher precision than naively expected.

\subsection{Scan Settings}

If the algorithm is used in a multidimensional scan of supersymmetric parameters, an example using SuSpect3 is in~\cite{Butter:2015fqa}, the $m_h$MSSM ensures that no spectrum calculation will
be performed for parameter sets incompatible with the measured central value of $m_h$. This leads to a reduction of the parameter 
space allowed for $A_t$. However such a calculation, due to the iteration on $A_t$, has a calculational overhead compared to an MSSM
spectrum calculation.

Given that the experimental uncertainty on the 
$m_h$ measurement is of the order of 0.1~GeV and additionally the theoretical
uncertainty is about~2~GeV, the following calculations have been performed with reduced precision: the RGE iterations are stopped when percent level convergence on $\overline{m}_A$ has been reached. All other convergence definitions remain unchanged. These are the standard settings used in Suspect.

To illustrate the calculational overhead the MSSM calculation is compared to the $m_h$MSSM calculation for s3. 
The number of RGE iterations is increased from four to six. The algorithm typically adds one additional iteration to each EWSB calculation on top
of the three for the standard algorithm, i.e., a total
of 27 iterations is necessary compared to 13 for the standard settings. 
The calculation of $A_t$ at the first two RGE iterations as a direct calculation of the approximate solution is not computationally intensive. 
When the fixed point method is used, for the first two RGE iterations, at the first EWSB iteration six and four calculations 
are necessary to converge on the fixed point to the required accuracy. For the subsequent EWSB iterations typically only 
one or two iterations on $A_t$ are necessary. For the last two RGE iterations for all EWSB iterations only
one or two $A_t$ calculations are necessary. For half of 
EWSB iterations, a single calculation of $A_t$ is sufficient. The reduction of the number of iterations on $A_t$ as  
the RGE and EWSB iterations progress illustrates the convergence of the algorithm.

\begin{table}[htb]
\begin{center}
    \caption{The result of the Higgs inversion algorithm  for the three definitions of EWSB input variables is shown. 
    s3 is the true point. The calculated $A_t$ and the resulting $m_h$ are listed for 
    all calculations.}
    \label{tab:HiggsInvScanSet}
    \begin{tabular}{llcccc}
         \toprule
         EWSB &        &  s1 & s2 & s3 & s4 \\
         \midrule
         $m^2_{H_d}$, $m^2_{H_u}$, sign($\mu$) & $A_t$ [GeV] & -5617.8   & -3795.0   & 3610.5  & 6085.9 \\
                                               & $m_h$ [GeV] &   125.012 &   125.012 & 125.012 & 125.012 \\
         \midrule
         $m_A^2(Q)$, $\mu$                     & $A_t$ [GeV] & -5606.9   & -3795.1   & 3610.7  & 6090.1 \\
                                               & $m_h$ [GeV] &   125.012 &   125.012 & 125.012 & 125.012 \\
         \midrule
         $m_A$, $\mu$                          & $A_t$ [GeV] & -5607.2   & -3794.7   & 3610.7  & 6089.9 \\
                                               & $m_h$ [GeV] &   125.012 &   125.012 & 125.012 & 125.012 \\
         \botrule
    \end{tabular}
\end{center}
\end{table}

In \cref{tab:HiggsInvScanSet} the results on $A_t$ and the calculated $m_h$ are shown for the reduced precision setting of the algorithm. The maximal difference between the calculated Higgs masses is less than 1~MeV, i.e., largely sufficient given the experimental error.

The two EWSB calculations with the $\mu$ parameter as input lead to almost identical results for all $A_t$ solutions.  The two EWSB calculations with $\mu$ use
$\overline{m}_{H_u}^2$ as variable to test convergence whereas the other EWSB calculation uses $\overline{m}_A^2$ to stop the RGE iterations. The calculation is performed with a fixed
convergence precision. The impact on the 
$m_h$ value for a change of $\overline{m}^2_{H_u}$ and $\overline{m}_A^2$ is not identical. Therefore 
the results for EWSB with the A boson mass, tree level or pole, can be different with respect to the calculation with
the Higgs mass parameters. 
For the true solution
s3, the maximal deviation for all EWSB variants is only two tenth of a permil\jl{,} at 0.7~GeV with respect to the true $A_t$ of the cliff point~\cref{tab:MhInvPoints}.

%

\subsection{Beyond the Cliff}

The cliff point is a favorable situation as four distinct solutions exist. A parameter set can lead to a situation where the local
minimum of the Higgs boson mass in the vicinity of $A_t=0$~GeV is larger than the $m_h$ parameter input. Alternatively 
the $m_h$
parameter could be higher than either one or both of the maxima of \cref{fig:mHAt}.

To test the validity of the algorithm beyond the proof of concept in the cliff point, the input
$m_h$ parameter was varied. The algorithm was slightly extended for $m_h$ values close to the maxima and the local minimum by a  bisection algorithm. 
To ensure a logical coherence of the results, the regions of validity for $A_t$ are defined for the four solutions:
\begin{itemize}
    \item s1: $-\infty$ to $A_t(m_h^{max}(A_t<0))$
    \item s2: $A_t(m_h^{max}(A_t<0))$ to $A_t(m_h^{min}(A_t\sim 0))$
    \item s3: $A_t(m_h^{min}(A_t\sim 0))$ to $A_t(m_h^{max}(A_t>0))$
    \item s4: $A_t(m_h^{max}(A_t>0))$ to $\infty$
\end{itemize}
If the $m_h$ parameter is out of reach, the closest $A_t$ in the region is used.

\begin{figure}[htb]
\begin{center}
\includegraphics[width=0.45\textwidth]{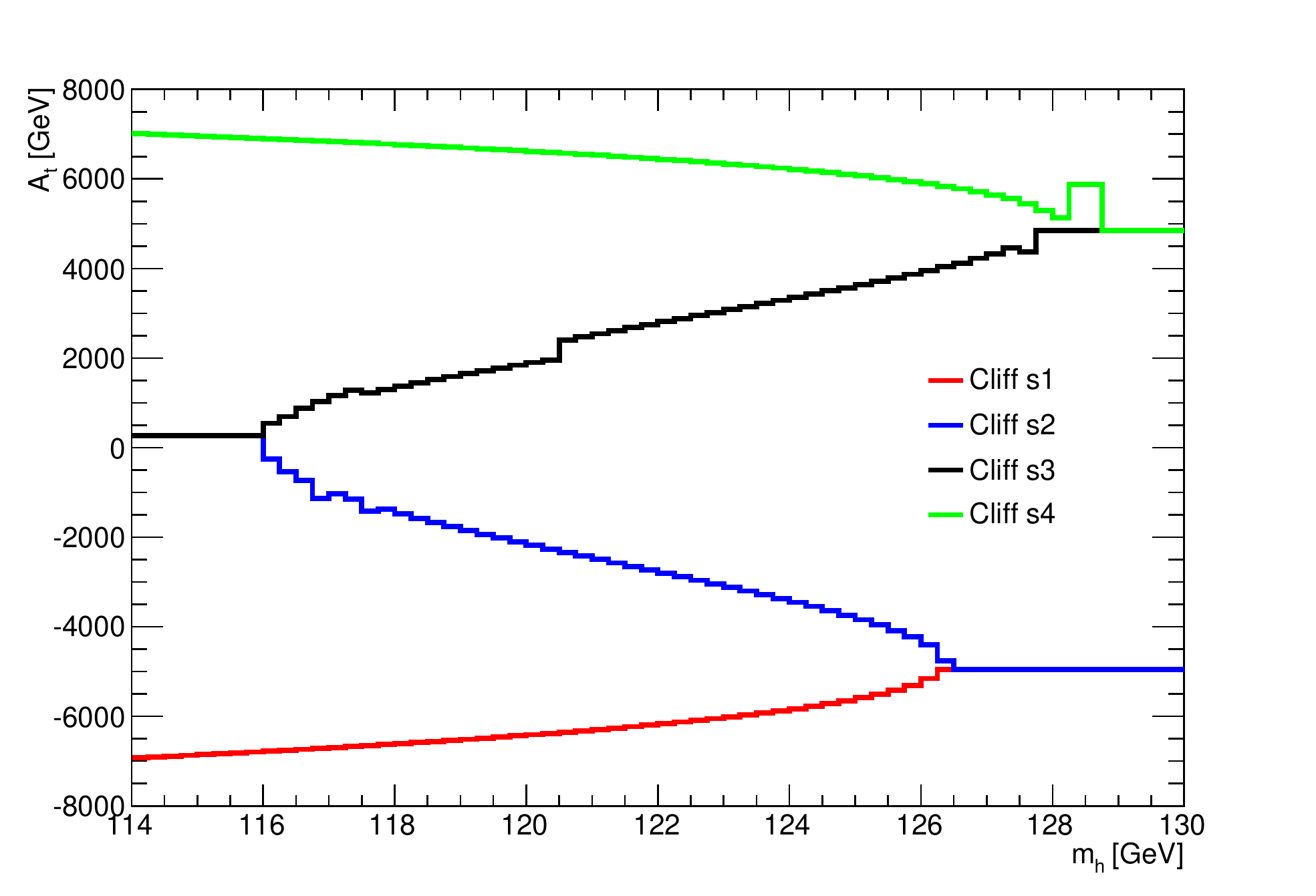}
\end{center}
\caption{The result of the determination of $A_t$ is shown as function of the chosen $m_h$. The determination is performed separately for the four solutions. The histograms are stacked from s1 (bottom) to s4 (top), the horizontal lines at constant $A_t$ at the transition between the regions show only the top color.
}
\label{fig:mhScan}
\end{figure} 

The result of a scan of $m_h$ values between 112~GeV and 128~GeV is shown in \cref{fig:mhScan}. The four solutions are correctly reconstructed
over the full range. For s2 and s3 the lowest parameter values cannot be reached, therefore the $A_t$ value of the local minimal mass is determined. A similar feature
is observed when the parameter is greater than the maximal $m_h$ at 2-loop precision shown in \cref{fig:mHAt}. In this case $A_t$ corresponding
to the maximal mass achievable for the parameters is determined, resulting in a line parallel to the x--axis.
The steps correspond to cases where the $\mu$ parameter and the CP-odd scalar Higgs boson mass have converged to a slightly different value which leads
to a step in $A_t$ and an offset in the predicted 
$m_h$, smaller than the systematic of~2~GeV typically used in scans of the parameter space. In total 256 inversions are shown of which only 3 have not converged correctly. These points are all located in regions where a larger $m_h$ than possible was requested (horizontal lines at $A_t \approx -4.95$~TeV and $A_t\approx 4.85$~TeV) and for two the algorithm returned an error for non convergence. Since the step size of the scan of 0.25~GeV is smaller than the error on the Higgs mass of about 2~GeV, posterior distributions of $A_t$ also give an additional handle on the outliers. 

\cref{fig:mhScan} shows $A_t$ as multi-valued function of $m_h$ covering the same parameter space as \cref{fig:mHAt}. 
This shows that the inversion, based on the exact calculation, works, validating the use of the approximate one--loop calculation as a first guess.  
However, there could be problematic parameter sets for which the guess points are not sufficiently close to the fixed points 
to ensure convergence. An alternative algorithm relying on the knowledge of $A_t(m_h^{max}(A_t<0)), A_t(m_h^{min}(A_t\sim 0))$ and $A_t(m_h^{max}(A_t>0))$, combined with global rather than local criteria, could then be used.


\subsection{Beyond the Fixed-Order Approximation}

Our inversion strategy is built upon fixed-order diagrammatic results, in particular upon the knowledge of the analytical form of the exact one--loop contribution to the Higgs mass. 
In more recent developments of MSSM Higgs mass radiative corrections, 
there are, however, configurations where an EFT approach is better
suited for a reliable estimate of the Higgs mass, like typically when considering an MSSM setup with (very) heavy scalars (see e.g. \cite{Slavich:2020zjv} for a recent review). In this case, if for instance a heavy stop sector is integrated out from the low-energy EFT, a direct analytical relation between $m_h$ and $A_t$ will be essentially lost. Nevertheless, an inversion procedure can still be carried out: 
The needed relation will reside now in the matching condition at the boundary scale between the light and heavy sectors for the quartic Higgs coupling, from which the multi-$A_t$ solutions can be retrieved analytically.

\subsection{Beyond the CP-conserving Case}
If complex phases are allowed for the MSSM parameters, the dependence of $m_h$ on $A_t$ will be in general significantly
modified due to loop induced CP-violating mixings among all the neutral Higgs states. The ensuing 
$3 \times 3$ mass matrix after identifying the Goldstone boson,  implies an 
analytical expression of $m_h^2$ given by a root of a cubic equation, quite different 
from that obtained from the diagonalization of \cref{eq:Ms}.
Our approach can be easily extended to this case in the (experimentally likely) configuration 
where the MSSM spectrum, and in particular the charged Higgs 
mass, are much heavier than the $125\mathrm{GeV}$ state. As well known, in this case 
the CP-violating mixing is essentially confined  to the two 
heavy neutral states sector, see e.g. \cite{Pilaftsis_1999}, and
the phenomenology (mass, couplings) of the lightest would-be CP-even state  becomes essentially the same as in the CP-conserving MSSM. In this limit, taking $A_t$ and $\mu$ complex (which is sufficient to account for the physically relevant phases, denoted $\phi_{A_t}$ and 
$\phi_{\mu}$), the functional dependence of $m_h$ on $\lvert A_t\rvert$ will be the same as that on $A_t$ in the CP-conserving case. The only difference is the reduction of some coefficients by cosine factors, (typically occurrences of
$(\mu A_t)^n$ become $(\left\lvert\mu \right\lvert \left\lvert A_t\right\lvert)^n \cos n (\phi_{\mu} +\phi_{A_t})$, otherwise even powers $A_t^{2n}$ become $\left\lvert 
A_t\right\lvert^{2n}$, etc.). Modulo these modifications in 
\cref{eq:AppRC,eq:mh4At},
our algorithm will work exactly as before, 
 to determine $\left\lvert A_t\right\lvert$ (and/or $-\left\lvert A_t\right\lvert$) solutions  from $m_h$, given  $\phi_{A_t}$ (and or $\phi_{A_t}\pm \pi$), a complex $\mu$, and the other parameters of the MSSM. 
Beyond this (relative decoupling) limit, the algorithm can still
provide an educated guess for $\pm \lvert A_t\rvert$ input, within a more general numerical algorithm for the CP-violating MSSM.

\section{Conclusion \label{sec:conclusion}}

The discovery and precise measurement of the Higgs boson motivates the redefinition of the 
MSSM as $m_h$MSSM. The algorithm presented in this 
paper replaces the trilinear coupling of the stop sector $A_t$ with
the measured $m_h$ as input parameter. 

The simplified expression of the radiative
corrections to $m_h$ and the exact full one--loop with the leading two--loop corrections
are assembled in an algorithm to calculate $A_t$. The algorithm has been applied to a benchmark point showing that the four solutions are determined with the expected
precision. A single parameter scan with 
$m_h$ as scan parameter shows that the $A_t$ dependence is
correctly reconstructed.

The general structure of the algorithm could also be applied to other parameters. Using
the $m_h$MSSM may speed up the exploration of supersymmetric
parameter space by ensuring the compatibility of all calculated spectra with the experimentally
measured $m_h$. Future work will center on applying the algorithm in multi-parameter scans.

\bmhead{Acknowledgments}
We would like to thank Steve Muanza for his collaboration and input at an early stage of this study, as well as the PESBLADE working group of the OCEVU Labex for useful discussions. We also acknowledge instructive exchanges with Pietro Slavich. This work benefited from partial support
 of the OCEVU
Labex (ANR-11-LABX-0060) and the A*MIDEX project
(ANR-11-IDEX-0001-02) funded by the “Investissements
d’Avenir” French government program managed by the
ANR.

\begin{appendices}
\crefalias{section}{appendix}
\clearpage
\begin{boldmath}
\section{The coefficients of \higgsmolar$(A_t)$ \label{app:higgsmolar}}
\end{boldmath} 
We give hereafter the explicit dependence of the coefficients appearing in \cref{eq:mh4At} on the various
$\pi_{ij}^{(..)}$, tadpoles and running mass parameters, as well as the relevant combinations of the latter in terms
of the MSSM parameters of the stop sector.

\begin{equation}
\begin{aligned}
C_0[A_t] = &c_s  (\pi_{11}^{(s)} - t_1^{(s)})(\pi_{22}^{(s)}-t_2^{(s)})  + 
     (m_h^2 + \pi_{11}^{(0)} - t_1^{(0)} - \overline{m}_{11}^2)\\ 
     &\cdot(m_h^2 + \pi_{22}^{(0)} - t_2^{(0)} - \overline{m}_{22}^2) 
     - (\pi_{12}^{(0)} - \overline{m}_{12}^2)^2,\\ 
   C_1[A_t] = &(\pi_{11}^{(s)} - t_1^{(s)}) (b_s (\pi_{22}^{(s)}-t_2^{(s)}) - c_s t_2^{(1s)}) +
   (\pi_{22}^{(1)}-t_2^{(1)}) \\ 
   &\cdot (m_h^2 + \pi_{11}^{(0)} - t_1^{(0)} - \overline{m}_{11}^2) + 2 \pi_{12}^{(1)} (\overline{m}_{12}^2-\pi_{12}^{(0)}) ,\\ 
   C_2[A_t] =   &(\pi_{11}^{(s)} - t_1^{(s)}) (a_s (\pi_{22}^{(s)}-t_2^{(s)}) - b_s t_2^{(1s)}) \\
   &+\pi_{22}^{(2)} (m_h^2 + \pi_{11}^{(0)} - t_1^{(0)} - \overline{m}_{11}^2) -(\pi_{12}^{(1)})^2 ,\\ 
   C_3[A_t] = &a_s ( t_1^{(s)} -\pi_{11}^{(s)}) t_2^{(1s)} ,\\ 
   R_0[A_t] =  &  (\pi_{11}^{(s)}  - t_1^{(s)}) (m_h^2 + \pi_{22}^{(0)} - t_2^{(0)} - \overline{m}_{22}^2) \\ 
                 &+ (\pi_{22}^{(s)}  - t_2^{(s)}) (m_h^2 + \pi_{11}^{(0)} - t_1^{(0)} - \overline{m}_{11}^2),\\ 
   R_1[A_t] =  &(t_1^{(s)}  -\pi_{11}^{(s)}) (t_2^{(1)} -\pi_{22}^{(1)})+ 
      ( t_1^{(0)}  - \pi_{11}^{(0)} + \overline{m}_{11}^2 -m_h^2 ) t_2^{(1s)} ,\\ 
     R_2[A_t] = &\pi_{22}^{(2)} (\pi_{11}^{(s)} - t_1^{(s)}) ,
\end{aligned}
\end{equation}
where the $\overline{m}_{ij}^2$ are given by 
\cref{eq:m112,eq:m222,eq:m122}.

Relying on the full one--loop results (and partly on the notations) of \cite{Pierce:1996zz},\footnote{with, however, an opposite sign convention for $\mu$ in
accord with SuSpect3~\cite{Djouadi:2002ze,Brooijmans:2012yi}
and denoting the couplings $\lambda_{s \tilde{u} \tilde{u}}$ and $\lambda_t$ of \cite{Pierce:1996zz} by $g_{s \tilde{t} \tilde{t}}$ and $y_t$.} we extract the relevant contributions of the self-energies and tadpoles: 

\begin{align}    
16 \pi^2 (t_1^{(0)}  - \pi_{11}^{(0)})=&\frac{3}{\sqrt{2}}  \frac{g_2 c_t s_t y_t \mu M^2}{c_\beta M_W} \ln\left( \frac{m_{\tilde{t}_1}}{m_{\tilde{t}_2}}\right) \nonumber \\
&-3 \sum_{i,j=1}^2 g_{s_1 \tilde{t}_i \tilde{t}_j}^2 B_0\left( m_{\tilde{t}_{i}},m_{\tilde{t}_{j}}\right) + \cdots, \label{eq:pidiffs1}\end{align}
\begin{align}
16 \pi^2 (t_1^{(s)}  -\pi_{11}^{(s)})&=  \frac{3}{\sqrt{2}}  \frac{g_2 c_t s_t y_t \mu}{c_\beta M_W}\left(1 - \ln\left( \frac{m_{\tilde{t}_1}m_{\tilde{t}_2}}{Q^{2}}\right)   \right), \label{eq:pidiffs2}\end{align}
\begin{align}
16 \pi^2(t_2^{(0)}-\pi_{22}^{(0)})= -3\Big(& \sum_{i=1}^2 (\Delta g_i)^2  B_0\left( m_{\tilde{t}_{i}},m_{\tilde{t}_{i}}\right) \nonumber \\
&+ 2 (\Delta g_{12})^2  B_0\left( m_{\tilde{t}_{1}},m_{\tilde{t}_{2}}\right)\Big) + \cdots, \end{align}
\begin{align}
16 \pi^2 (t_2^{(s)} - \pi_{22}^{(s)})&= 0, \label{eq:pidiffs4}\end{align}
\begin{align}
16 \pi^2(t_2^{(1)}- \pi_{22}^{(1)})=&  \frac{3}{\sqrt{2}}  \frac{g_2 y_t c_t s_t M^2}{s_\beta M_W} \ln\left( \frac{m_{\tilde{t}_2}}{m_{\tilde{t}_1}} \right)
 \nonumber\\
& - 6 \sqrt{2} y_t \Big( c_t s_t\left(\Delta g_1 B_0\left( m_{\tilde{t}_{1}},m_{\tilde{t}_{1}}\right) - \Delta g_2 B_0\left( m_{\tilde{t}_{2}},m_{\tilde{t}_{2}}\right)\right)
\nonumber \\
 &~~~~~~~~~~~~~+ (c_t^2 - s_t^2) \Delta g_{12} B_0\left( m_{\tilde{t}_{1}},m_{\tilde{t}_{2}}\right) \Big),  \label{eq:pidiffs5}\end{align}
\begin{align}
16 \pi^2 t_2^{(1s)}&=  \frac{3}{\sqrt{2}}  \frac{g_2 y_t c_t s_t}{s_\beta M_W}\left(\ln\left( \frac{m_{\tilde{t}_1}m_{\tilde{t}_2}}{Q^{2}}\right) -1  \right), \label{eq:pidiffs6}\end{align}
\begin{align}
16 \pi^2 \pi_{22}^{(2)}= 3 y_t^2 \Big(& 2 c_t^2 s_t^2 \left( B_0\left( m_{\tilde{t}_{1}},m_{\tilde{t}_{1}}\right) + B_0\left( m_{\tilde{t}_{2}},m_{\tilde{t}_{2}}\right) \right)
\nonumber \\
&+  (c_t^2 -s_t^2)^2 B_0\left( m_{\tilde{t}_{1}},m_{\tilde{t}_{2}}\right) \Big), \\
16 \pi^2 \pi_{12}^{(0)} =  3 \Big(&  g_{s_1 \tilde{t}_1 \tilde{t}_1} \Delta g_{1} B_0\left( m_{\tilde{t}_{1}},m_{\tilde{t}_{1}}\right) +  g_{s_1 \tilde{t}_2 \tilde{t}_2} \Delta g_{2} B_0\left( m_{\tilde{t}_{2}},m_{\tilde{t}_{2}}\right)  \nonumber\\
&+  g_{s_1 \tilde{t}_1 \tilde{t}_2} \Delta g_{12} B_0\left( m_{\tilde{t}_{1}},m_{\tilde{t}_{2}}\right) \Big) + \cdots,~~~~~~~~~~~~~~~~~~~~~  \\
16 \pi^2 \pi_{12}^{(1)}= \frac{3}{\sqrt{2}}  y_t \Big(& 2 c_t s_t \left(g_{s_1 \tilde{t}_1 \tilde{t}_1} B_0\left( m_{\tilde{t}_{1}},m_{\tilde{t}_{1}}\right) -g_{s_1 \tilde{t}_2 \tilde{t}_2} B_0\left( m_{\tilde{t}_{2}},m_{\tilde{t}_{2}}\right) \right) \nonumber \\
&+2 (c_t^2 - s_t^2) g_{s_1 \tilde{t}_1 \tilde{t}_2} B_0\left( m_{\tilde{t}_{1}},m_{\tilde{t}_{2}}\right) \Big) 
\end{align}
where we have defined
\begin{equation}
\begin{aligned}
\Delta g_{1} &\equiv  c_t^2 g_{s_2 \tilde{t}_L \tilde{t}_L} + s_t^2 g_{s_2 \tilde{t}_R \tilde{t}_R}, \\
\Delta g_{2} &\equiv s_t^2 g_{s_2 \tilde{t}_L \tilde{t}_L} + c_t^2 g_{s_2 \tilde{t}_R \tilde{t}_R}, \\
\Delta g_{12} &\equiv c_t s_t (g_{s_2 \tilde{t}_R \tilde{t}_R}-g_{s_2 \tilde{t}_L \tilde{t}_L}).
\end{aligned}
\end{equation}

$s_\beta$ and $c_\beta$ stand for $\sin \beta$ and $\cos \beta$, and all other quantities have been defined previously, see also \cref{eq:mstops,eq:asetc,eq:gs2}. The ellipses indicate contributions from the Higgs/Higgsino, gauge/gaugino and other scalar/fermion sectors, that do not have a direct dependence on $A_t$.
The couplings $g_{s_1 \tilde{t}_i \tilde{t}_j}$ are related to $g_{s_1{\tilde{t}_{L,R}}{\tilde{t}_{L,R}}}, g_{s_1{\tilde{t}_L}{\tilde{t}_R}}$ as in \cref{eq:gs2}, but do not depend explicitly on $A_t$.
All parameters appearing in the above expressions are  understood to be running in the $\dr$ scheme, and $Q$ is the corresponding renormalization scale. The dependence on 
$p^2$, cf. \cref{eq:Ms}, taken here at $p^2=m_h^2$, is in the
$B_0$ functions with the shorthand notation $B_0(m_1,m_2) \equiv B_0(m_h, m_1,m_2)$, see also \cref{footnote:real}.

Note finally that the vanishing of $t_2^{(s)} - \pi_{22}^{(s)}$, cf. \cref{eq:pidiffs4}, together with the fact that
$M^2$ appears exclusively with $\ln\left( \frac{m_{\tilde{t}_1}}{m_{\tilde{t}_2}}\right)$, \cref{eq:pidiffs1,eq:pidiffs5}, and
$\sqrt{a_s A_{t}^{2}+b_s A_{t} +c_s}$ exclusively with $\ln\left( \frac{m_{\tilde{t}_1}m_{\tilde{t}_2}}{Q^{2}}\right)$, \cref{eq:pidiffs2,eq:pidiffs6}, are a direct consequence of the cancellation of the quadratic divergences before renormalization as expected in softly-broken SUSY. Indeed, 
in this case the stop sector contributions from the $A_0$ function can occur only in the combination $A_0(m_{\tilde{t}_1}) - A_0(m_{\tilde{t}_2})$.\\


\section{Convergence Criterion \label{app:convergence}}
A fixed point $x_{\rm FP}$ of a function $F$, satisfying $F(x_{\rm FP})=x_{\rm FP}$, can be determined iteratively
as the limit of a sequence defined by ${x}_{i+1} =  F(x_i)$ and an initial guess $x_0$, only if the fixed point
is attractive, i.e. $-1 \leq F'(x_{\rm FP}) \leq 1$. We sketch here how to proceed in the more general cases of
non-attractive fixed points:

Define
\begin{equation}
F_\tau(x)= \frac{ (\tau-1) x + F(x)}{\tau}, 
\rm{\; with} \, \tau \neq 0 , \label{eq:Ftau}
\end{equation}
and
\begin{equation}
F(x) - F(x_0) = (x-x_0) K(x,x_0) \label{eq:F-K}
\end{equation}
with
\begin{equation}
K(x_0,x_0) = F'(x)_{\lvert x=x_0}.
\end{equation}
Let us consider two distinct sequences given by
\begin{equation}
{x}_{i+1} =  F_\tau(x_i) \text{~and~}  {y}_{i+1} =  F_\tau(y_i), \ (\text{with} \ x_0 \neq y_0), \label{eq:itert}
\end{equation}
and study the variation $F_\tau(x_{n}) - F_\tau(y_{n})$
after $n$ iterations. One finds straightforwardly upon repeated use of \cref{eq:Ftau,eq:F-K,eq:itert}:

\begin{eqnarray}
F_\tau(x_{n}) - F_\tau(y_{n}) &=& (x_{n}-y_{n}) 
\left(1 + \frac{K(x_{n},y_n) - 1}{\tau}\right), \nonumber \\
&=& \left(F_\tau(x_{n-1}) - F_\tau(y_{n-1}) \right) 
\left(1 + \frac{K(x_{n},y_n) - 1}{\tau}\right), \nonumber \\
&\vdots & \nonumber \\
F_\tau(x_{n}) - F_\tau(y_{n}) &=& (x_{0}-y_{0}) 
\prod_{i=0}^{n}\left(1 + \frac{K(x_{i},y_i) - 1}{\tau}\right) . \label{eq:lipschitz}
\end{eqnarray}
It follows from Eq.~(\ref{eq:lipschitz}) that 
\begin{equation}
\text{\parbox{.80\textwidth}{\sl{ choosing $\tau$ such that}
$\left\vert 1 + \frac{K(x_{i},y_i) - 1}{\tau}\right\rvert$ {\sl is sufficiently smaller than $1$ for a sufficiently large set of $i$'s between $0$ and $n$,}}}
\label{eq:conv-cond}
\end{equation}
leads to
\begin{equation}
\left\lvert\prod_{i=0}^{n}\left(1 + \frac{K(x_{i},y_i) - 1}{\tau}\right) \right\rvert \ll 1 \label{eq:conv-cond1},
\end{equation}
and the two sequences $x_i$ and $y_i$ converge towards each other aftern $n$ iterations within a prescribed numerical 
precision. In particular, choosing $y_0=x_{\rm FP}$ yields a constant sequence
$\{y_i\}=\{x_{\rm FP}\}$, and $x_i$ converges to the fixed point $x_{\rm FP}$. This completes the proof that by using the auxiliary function
$F_\tau$, one can in principle always optimize the choice of $\tau$ depending on the variations of $F$ 
, so as to converge to $x_{\rm FP}$ for a given initial guess value $x_0$, even when dealing with non-attractive fixed points of $F$. It should be stressed that $K(x_i,y_i)$ is not always a very good estimate of the local variation of $F$ since the two sequences $\{x_i\}$ and $\{y_i\}$ do not necessarily start off close to each other. 
Nonetheless, it will become so after a few iterations 
if $\tau$
satisfies the criterion (\ref{eq:conv-cond}). In particular,  for the constant sequence $\{y_i\}\equiv \{x_{\rm FP}\}$, the successive values of $K$ correspond to variations with respect to the same reference point $x_{\rm FP}$ which is of course not yet known. However, if the initial guess point $x_0$ is sufficiently close to $x_{\rm FP}$, one can in practice use
$K(x_i,x_{i+1})$ instead of $K(x_i,x_{\rm FP})$ to optimize $\tau$ piece-wise.
With this in mind, we can apply the above procedure to the case of $\linearFP$, where the quantity $\left(1 + \frac{K(x_{i},x_{i+1}) - 1}{\tau}\right)$, with $K(x_{i},x_{i+1}))=\Delta \linearFP/\Delta x $ being an estimate of $\linearFP'(x_{\rm FP})$ for $x_{i}$ sufficiently close to $x_{\rm FP}$, is clearly a discretized version of $\linearFP'_\tau(x_{i})$ as given by \cref{eq:dt}. 
A simple numerical algorithm can thus be devised, based on
\cref{eq:algotau} as described at the end of \Cref{subsec:FullOneLoopFP}.

\end{appendices}

\bibliography{HiggsInversion}

\end{document}